\documentclass{article}
\usepackage[utf8]{inputenc}
\usepackage{amsmath, amsfonts, amsthm}
\usepackage{graphicx}
\usepackage{svg}
\usepackage{caption}
\usepackage{tabularx}
\usepackage{url}
\usepackage{subcaption}

\newcommand\Tstrut{\rule{0pt}{2.6ex}}         

\newcommand{\rs}[1]{{\color{black} #1}}

\usepackage[a4paper,
            bindingoffset=0.2in,
            left=0.5in,
            right=0.5in,
            top=0.5in,
            bottom=0.5in,
            footskip=.25in]{geometry}

\usepackage{setspace}
\usepackage{natbib}
\usepackage{appendix}
\usepackage{cases}
\usepackage{empheq}
\usepackage{authblk}

\title{Modeling the extracellular matrix in cell migration and morphogenesis: A guide for the curious biologist}
\author[1, $\dagger$]{Rebecca M. Crossley}
\author[1, $\dagger$]{Samuel Johnson}
\author[2, $\dagger$]{Erika Tsingos}
\author[3]{Zoe Bell}
\author[4,5]{Massimiliano Berardi}
\author[6]{Margherita Botticelli}
\author[7]{Quirine J.S. Braat}
\author[8,9]{John Metzcar}
\author[10]{Marco Ruscone}
\author[1]{Yuan Yin}
\author[11, $*$]{Robyn Shuttleworth}

\affil[1]{Wolfson Centre for Mathematical Biology, Mathematical Institute, University of Oxford, UK}
\affil[2]{Computational Developmental Biology Group, Institute of Biodynamics and Biocomplexity, Utrecht University, Utrecht, NL}
\affil[3]{Northern Institute for Cancer Research, Newcastle University, UK}
\affil[4]{LaserLab, department of Physics and Astronomy, Vrije Universiteit Amsterdam, Amsterdam, NL}
\affil[5]{Optics11 life, Amsterdam, NL}
\affil[6]{Watson School of Mathematics, University of Birmingham, UK}
\affil[7]{Department of Applied Physics and Science Education, Eindhoven University of Technology, NL}
\affil[8]{Department of Intelligent Systems Engineering, Indiana University. Bloomington, IN, USA}
\affil[9]{Department of Informatics, Indiana University. Bloomington, IN, USA}
\affil[10]{Institut Curie, Université PSL, F-75005 Paris, France}
\affil[11]{Altos Labs, Redwood City, CA, USA}
\affil[$\dagger$]{\textbf{These authors contributed equally to this work and share first authorship.}}
\affil[$*$]{Corresponding author: \tt rshuttleworth@altoslabs.com}

\date{}

\begin{document}
\onecolumn

\maketitle
\setlength{\marginparwidth}{10pt}






\begin{abstract}

The extracellular matrix (ECM) is a highly complex \rs{structure} through which biochemical and mechanical signals are transmitted. 
In processes of cell migration, the ECM also acts as a scaffold, providing structural support to cells as well as points of potential attachment. 
Although the ECM is a well-studied structure, its role in many biological processes remains difficult to investigate comprehensively due to its complexity and structural variation within an organism. 
In tandem with experiments, \rs{mathematical models are helpful in refining and testing hypotheses, generating predictions, and exploring conditions outside the scope of experiments.} 
Such models can be combined and calibrated with \textit{in vivo} and \textit{in vitro} data to \rs{identify critical} cell-ECM interactions that drive developmental and homeostatic processes, or the progression of diseases. 
In this review, we focus on mathematical and computational models of the ECM in processes such as cell migration including cancer metastasis, and in tissue structure and morphogenesis.
By highlighting the predictive power of these models, we aim to help bridge the gap between experimental and computational approaches to studying the ECM and to provide guidance on selecting an appropriate model framework to complement corresponding experimental studies.
\newline
\noindent\hrulefill


\end{abstract}

\section{Introduction}
The extracellular matrix (ECM) \rs{encompasses all biological components outside of cells. It provides biochemical cues and acts} as a physical scaffold for cells and tissues, facilitating cell-cell and cell-microenvironment communication \citep{bowers2010extracellular}. 
The ECM is present within all tissues and organs and its composition is highly diverse, varying with the type and location of tissue in an organism \citep{karamanos2021guide}.
This structural variation is further amplified by cell-mediated ECM remodeling. For example, cells can realign or crosslink ECM components, thereby changing matrix stiffness \citep{Karsdal2013}, or directly degrade the ECM to expedite invasion \citep{artym2009ecm}. 
Additionally, the ECM provides essential biochemical and biomechanical cues for cell differentiation, tissue morphogenesis, and homeostasis \citep{yamada2019extracellular, HeLeeJiang_2022}. 
The mechanical effects of the ECM are particularly important in facilitating cell movement during embryonic development, wound healing, and in the progression of diseases such as fibrosis and cancer \citep{walma2020extracellular,Diller_2022,Dzobo2023}. For example, an increase in ECM stiffness is associated with the progression of \rs{tumor malignancy} \citep{guo2022stiffness}.

\rs{The highly complex nature of the ECM} means that it remains difficult to fully quantify \rs{and understand} the role of all components through experiments alone. Mathematical and computational models, therefore, are well-poised to provide further insights 
via rapid simulation of complex biological processes involving the ECM at a lower cost than corresponding experiments.

\rs{The aim of this review is to highlight the various approaches used to model biological systems in which the ECM plays a major role. 
We begin by briefly reviewing ECM structure and function, as well as experimental and modeling approaches.
We then give an in-depth overview of theoretical work that includes model representations of the ECM, focusing on models of (1) cell migration (including wound healing and cancer invasion) and (2) tissue structure and morphology (including morphogenesis). 
We selectively highlight models that explicitly represent the ECM and cover a breadth of techniques. Each section begins with a short overview of the underlying biological processes and reviews the different types of models that have been developed in each context, highlighting key results. 
Finally, we provide guidance on choosing the most appropriate modeling framework for a given scientific inquiry and comment on the current open challenges surrounding modeling of the ECM. We also list potential limitations of these models and highlight the importance of collaborations to overcome these difficulties in further uncovering the role of the ECM in biological systems.}

\subsection{ECM composition and properties} 

 \rs{The average mammalian tissue consists of over 150 ECM and ECM-associated proteins, and several hundred proteins are catalogued in the matrisome database \citep{shao2020matrisomedb}, testifying to the enormous quantity of molecular components and the diversity of ECM composition of different tissues within an organism.} 
This diversity is amplified by continuous ECM remodeling through both enzymatic and non-enzymatic interactions. \rs{Remodeling not only alters ECM composition, but also the three-dimensional (3D) structural organization of its molecular components, which determines the ECM's physical properties \citep{Chaudhuri2020}.} 
The composition and structure of the ECM changes most greatly during times of stress, for example during aging, tissue wounding, and tumor development.

There are two major forms of ECM \textit{in vivo}: interstitial ECM, and basement membrane (BM) ECM \citep{walma2020extracellular}. 
Interstitial ECM \rs{fills spaces between organs}. It is rich in fibrous proteins and proteoglycans that form 3D structures, whilst BMs form two-dimensional (2D) sheet-like ECM, \rs{that line organ boundaries}. 
A major structural component of both interstitial ECM and BMs is fibroblast-secreted collagen. 
There are currently 28 known types of collagens with collagen~I being the most common in interstitial ECM, and collagen~IV being the most abundant within BMs \citep{RicardBlum2011}. 
In interstitial ECM, collagen~I fibers undergo crosslinking to form networks, either with themselves or with other fibrous proteins, such as fibronectin and elastin, in processes facilitated by the enzyme lysyl oxidase (LOX) \citep{Sun2021}.
Of the other fibrous proteins, fibronectin facilitates the adherence of cells to the ECM, aiding their migration by providing new sites of potential adhesion \citep{Parisi2020}. Elastin, on the other hand, confers resilience to plastic deformation and is found most frequently in blood vessels, skin, and lung tissue \citep{Elastin_2016}.

In the BM, the main structural scaffold consists of two interconnected polymeric networks of laminin and collagen~IV.  \rs{Including splice variants, there are 16 known laminin complexes in humans \citep{mckee2021organization}.}
The laminin and collagen~IV networks do not interact directly but are crosslinked by other macromolecules such as nidogen and perlecan \citep{topfer2023basement}. 

\begin{figure}[ht]
    \centering
    \includegraphics[width=17cm]{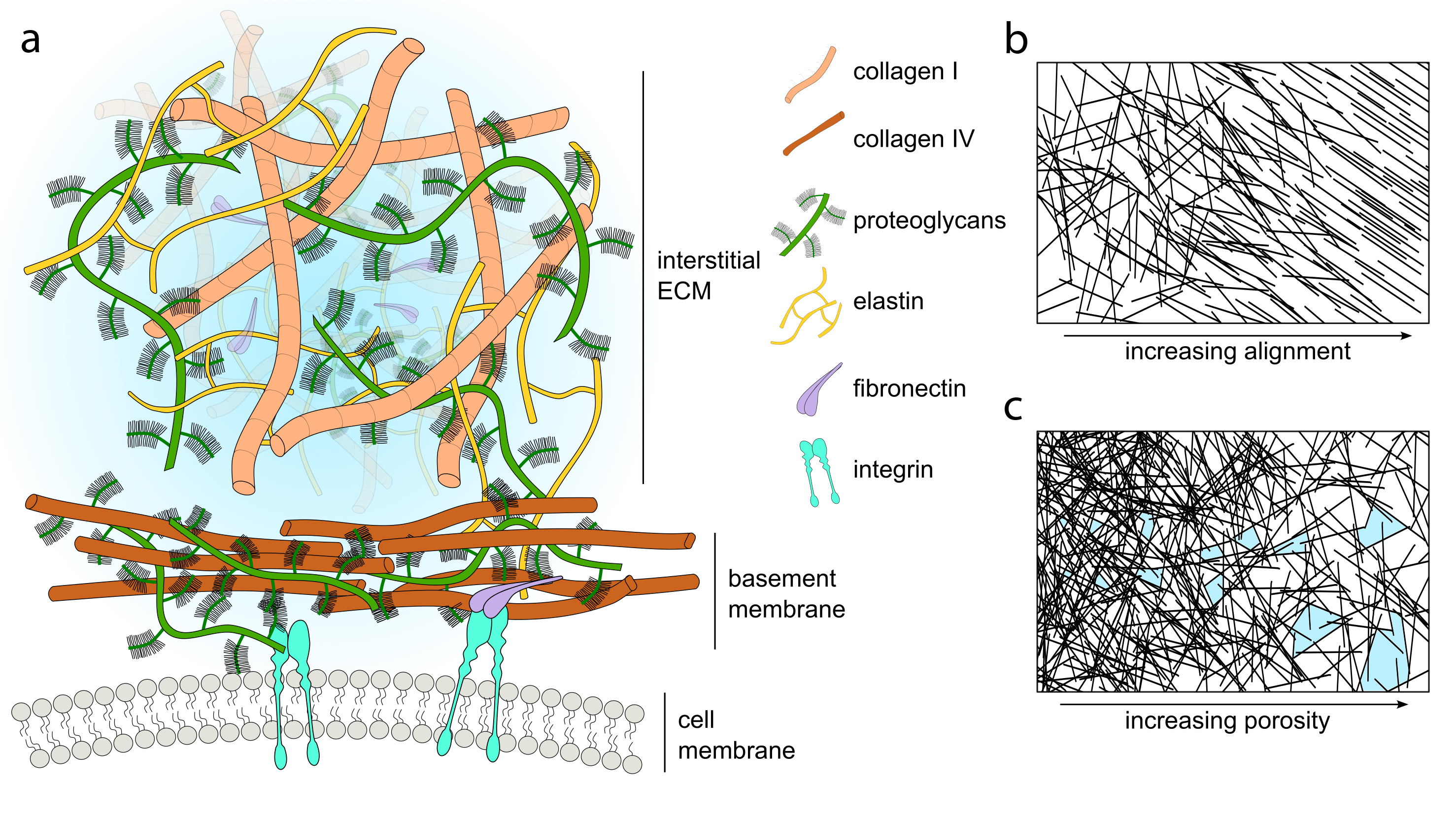}
    \caption{\rs{(a) Schematic representation of the major ECM compartments and their components. Macromolecular components of the ECM (size in the order of hundreds of nanometers) are smaller than cells (size in the order of tens of micrometers). However, many components such as laminin and collagen polymerize to form extended sheets or fibers (size order ranging from tens of micrometers to a few millimeters). For visual clarity, we do not represent small molecules such as cytokines, enzymes, or soluble factors. (b) Schematic representation of ECM fiber crosslinking and alignment. (c) Schematic representation of the change in ECM pore size with respect to ECM fiber density; pores highlighted in light blue.}}
    \label{fig:ECM_schematic}
\end{figure}

The ECM also contains a number of macromolecules, such as proteoglycans (PGs) and glycosaminoglycans (GAGs) \citep{frantz2010extracellular}. 
%
Additionally, the ECM stores proteases, such as matrix-degrading enzymes (MDEs) that target specific components of the ECM \citep{Parsons1997, Brinck2002}. 
These proteases play a key role in many biological processes, for example during wound healing, where matrix metalloproteinases (MMPs) degrade damaged collagen fibers, creating space for cells to migrate towards the site of wounding for tissue repair \citep{Kandhwal2022}.
Similarly, MMPs also play a significant role during tumor progression by degrading and remodeling the tumor microenvironment to facilitate cancer cell migration through the ECM \citep{Visse2003}.
\rs{A schematic highlighting some of the main components of ECM is shown in Figure \ref{fig:ECM_schematic}a, whilst Table \ref{tab:ecmfunc} lists the major ECM components and their functions.}

\rs{ With such a rich and diverse collection of components, and different ways of arranging those components into 3D architectures, it is natural that the biochemical and biomechanical properties of the ECM can vary greatly between tissues. 
%
%
Cells can interact with ECM via various receptors, such as integrins, discoidin domain receptors, and syndecans \citep{muncie2018physical, karamanos2021guide}. Cell-ECM interactions, in turn, modulate a number of intracellular signaling pathways, which results in changes to migratory behavior, proliferation, and adhesion in cell populations \citep{muncie2018physical}. PGs facilitate signaling between cells and their environment. For example, PGs can bind growth factors, cytokines, and morphogens, thus modulating their availability \citep{muncie2018physical, Barkovskaya2020}. Additionally, PGs and in particular GAGs such as hyaluronic acid, can bind large amounts of water, and thereby modulate tissue hydration \citep{karamanos2021guide}. Hydration impacts biochemical properties such as osmotic balance and the speed of molecular diffusion, and biomechanical properties such as porosity or the stiffness of other ECM components. For example, collagen becomes softer with increasing hydration \citep{andriotis2018collagen}.}

\rs{
The mechanical properties of the ECM are determined by its composition and structural organization at different length scales, ranging from the stiffness of molecular bonds at the nanoscale to protein fiber entanglement at the microscale \citep{Chaudhuri2020}. ECM biomechanical properties are often complex and time-dependent; important ECM material characteristics include stiffness, elasticity, plasticity, and viscosity. All of these properties can affect cell behavior, including alterations in cell spreading, proliferation, matrix deposition, and cell migration.

Perhaps the best-studied ECM mechanical property is stiffness. Stiffness describes a material's ability to resist deformation to an applied force. A simple example is a linear spring, whose spring constant determines its stiffness. More complex materials can have non-linear stiffness, time-dependent stiffness, or stiffness dependent on the rate and direction of force application. All of these responses can occur in different ECMs \citep{Chaudhuri2020}.
ECM stiffness depends primarily on the organization of collagen fibrils, with stiff matrices having a high density of collagen and fiber crosslinking \citep{Cox_2011}. Conversely, softer ECM has a less dense collagen network with a lower occurrence of fiber crosslinking (Figure \ref{fig:ECM_schematic}b). Matrix stiffness has consequences for many cellular processes. Membrane-bound mechanosensitive ion channels such as Piezo1 \citep{CHEN2018}, are over-expressed in stiff ECM and can cause uncontrolled cell proliferation in cancerous tissue. Additionally, stiff ECM tends to have reduced pore sizes (Figure \ref{fig:ECM_schematic}c), limiting cell migration in the absence of any proteolytic processes \citep{He_2022}. Each tissue has an optimal matrix stiffness, for example, load-bearing tissue such as bone and cartilage are very stiff in comparison to internal organs such as the liver and kidney \citep{Handorf_2015, Chaudhuri2020}. However, during aging and disease, mechanical homeostasis of the ECM is disrupted and matrix stiffness is altered, either increasing in stiffness through the increased deposition of collagen and increases in fiber crosslinking \citep{Rahman_2020}, or decreasing in stiffness due to an increase in MMP secretion that results in high levels of ECM degradation \citep{Wang_2023}.}

\rs{Elasticity refers to a material's ability to return to its original shape after an applied force has been removed. 
Crimped collagen fibers, elastic proteins such as elastin, and reversible swelling due to water-binding pores confer elasticity to the ECM \citep{Elastin_2016, karamanos2021guide, Chaudhuri2020}. 
Conversely, plasticity is the ability to retain a deformed shape after an applied force is removed. 
Intermolecular crosslinking can tune the balance between elasticity and plasticity in the ECM. Weaker non-covalent crosslinks enable flexible macromolecules to permanently rearrange, a feature of plasticity, whereas stronger or denser fiber crosslinking prevent deformation and confer elastic properties \citep{Chaudhuri2020}.
Viscosity refers to time-dependent features of the mechanical response. Everyday examples are water (a low-viscosity fluid) and honey (a high-viscosity fluid). A high density of water-binding molecules such as PGs and GAGs can increase ECM viscosity \citep{karamanos2021guide, Chaudhuri2020}. Additionally, weak non-bonded interactions between macromolecules and crosslinks that dynamically unbind upon force loading can also lead to ECM viscosity \citep{Chaudhuri2020}.
}

\rs{For a more in-depth review of the biochemical and biomechanical properties of the ECM, we refer the reader to excellent reviews elsewhere \citep{frantz2010extracellular, muncie2018physical, Chaudhuri2020, karamanos2021guide}.}

\subsection{Experimental approaches and data collection}\label{sec:experiment}
Experiments can be designed to characterize several features of the ECM and its components, namely its chemical composition, structural arrangement, and mechanical response to environmental stresses and strains. 
In such studies, an emphasis is often placed on characterizing the topography, viscoelastic properties such as poroelasticity \citep{Ehret2017}), and ligand presentation of the ECM \citep{Young2016}.

To explore the chemical composition of the ECM, commonly utilized methods include mass spectrometry, alongside immunostaining and fluorescence microscopy to characterize the ECM proteome \citep{Naba2017,Permlid2021}.
Spectroscopic methods such as Fourier transform infrared spectroscopy \citep{Chrabaszcz2020}, and Raman spectroscopy \citep{Ettema2022}, allow for the collection of spatially resolved spectra containing the vibrational fingerprint of the ECM.
These, in turn, permit the location and quantification of any molecular species present in a sample, as well as any variations in ECM structure \citep{Tiwari2020}. As signal detection in these methods relies on changes in vibrational states of molecules, neither require stains or dyes (in contrast with fluorescence microscopy). This is particularly useful when analyzing evolving structures non-invasively, as in the case the zona pellucida of oocytes during maturation \citep{Jimenez2019}.

The structural characterization of tissues is typically performed via a combination of microscopy methods. 
Common approaches include widefield and point scanning fluorescence methods such as confocal microscopy, a method that has recently been used to measure long-range cell-cell mechanical interactions via the ECM \citep{Nahum2023}. 
Other fluorescence-based super-resolution techniques exploit sophisticated illumination systems, for example stimulated emission depletion microscopy (STED), or the stochastic activation of fluorophores to improve on resolution limits \citep{Poole2021}. 
Further methods worth noting are based on higher harmonic generation. 
In this approach to ECM characterization, contrast in images arises from the intrinsic optical properties of the sample which avoids the need for staining and other methods of sample preparation that are otherwise necessary \citep{VanHuizen2020}. 
Due to their performance in imaging collagen fibers, second harmonic generation microscopes are especially well-suited to the quantification of fibrous structures \citep{Woessner2022}. 
If observations on smaller length scales are required, scanning or transmission electron microscopy offer viable solutions \citep{Leonard2018}.

The mechanical properties of tissues have been investigated across several different scales using a variety of experimental techniques. 
At the nanoscale and microscale, the most common characterization method is indentation, conducted using an atomic force microscope \citep{Plodinec2012}, or an instrumented nanoindenter \citep{Martinez-Vidal2023}. 
The selection of indenters of different dimensions affords great flexibility in quantifying both elastic and viscous properties, as well as providing the capacity to collect correlative measurements by pairing the system with brightfield or fluorescence microscopy. As a measurement typically consists of programmatically deflecting a cantilever over a set region, starting from an out-of-contact condition, it is also possible to extrapolate the sample topography. Other notable approaches include optical and magnetic tweezers \citep{Lehmann2020}, as well as fluorescence resonance energy transfer (FRET) biosensors \citep{Arnoldini2017}.

Common methods to investigate the mechanical properties of cells within the ECM are tensile testing \citep{Yang2015}, and rheometry \citep{Vos2020}. 
The former is more common when investigating tissues like skin and bone, and the latter is more convenient when studying reconstituted systems or samples that are particularly soft. 
A less common approach is that of optical coherence elastography, a method that pairs an optical coherence tomography system with mechanical loading of the tissue \citep{Kennedy2014}. 
Typically, this is realized by means of compression \citep{Hepburn2020}, suction \citep{Berardi2023}, or acoustic radiation force \citep{Li2020}. 
By virtue of light-tissue interactions, optical coherence elastography allows the rapid collection of 3D maps of scattering and mechanical contrast, spanning hundreds of microns in each direction. 
A relatively recent innovation is Brillouin microscopy \citep{prevedel2019brillouin}, a non-contact mechanical imaging method based on acoustically induced inelastic light scattering. Example applications include measuring mechanical properties of diseased cornea \citep{Shao2019} and developing zebrafish tissue \citep{Bevilacqua2019}.




\subsection{Modeling approaches}
\rs{Mathematical models can be used to describe a wide range of biological processes and explore hypotheses, providing a complementary means of investigation to laboratory experiments.} 
In general, experimental design should consider the equipment required to answer the research question at hand, and the same is true for choosing an appropriate mathematical modeling framework. 
However, it is important to stress that there is no `best' modeling approach for a given problem, and what is most suitable for a given study depends on the scientific question at hand and any prior knowledge of a system. 
In this section, we introduce some common approaches for mathematical modeling of the ECM, before detailing specific biological applications of these models in Sections~\ref{cellmig}~and~\ref{tissue-morph}.

\begin{figure}[ht]
    \centering
    \includegraphics[width=\textwidth]{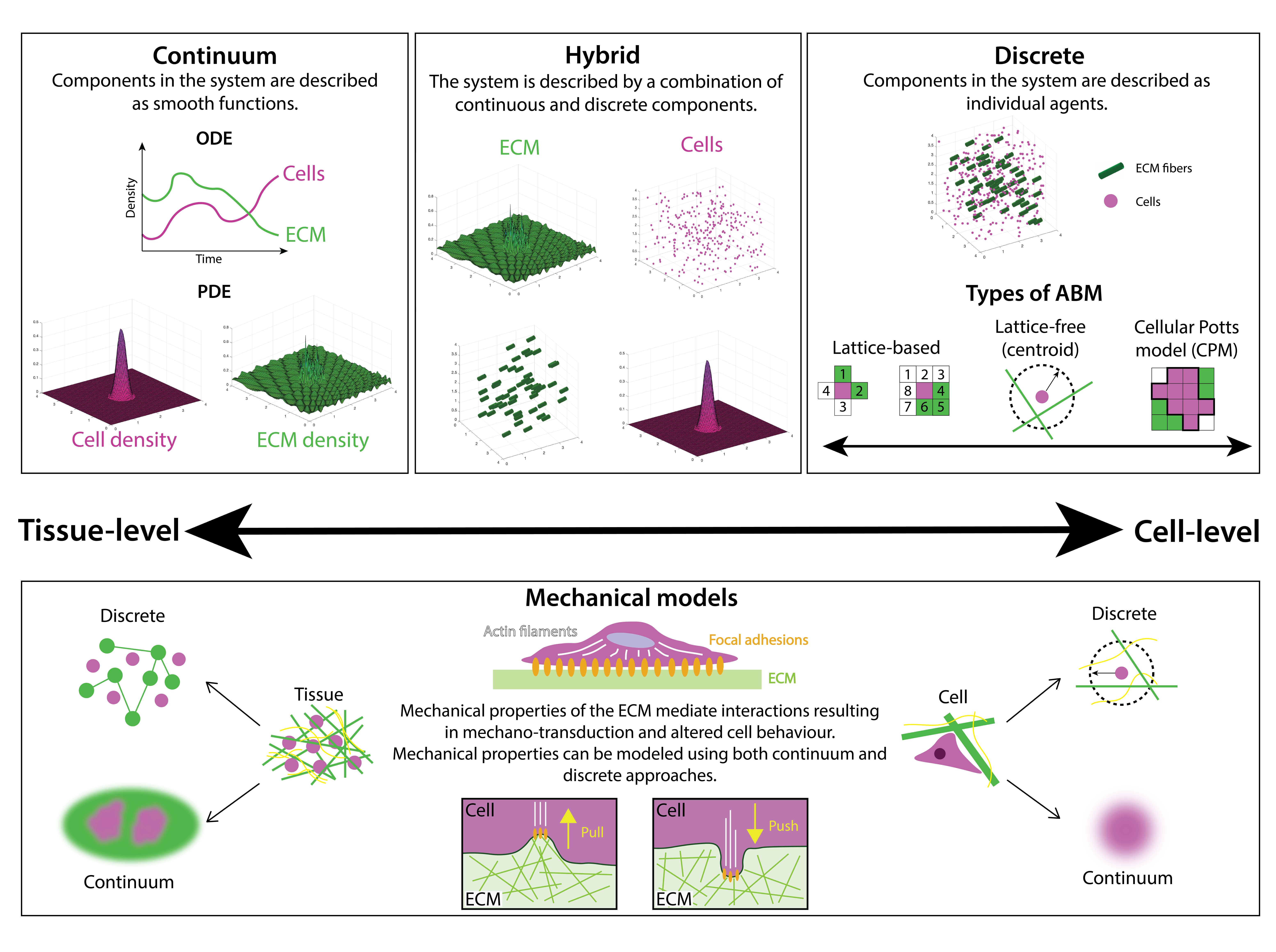}
    \caption{
    There are many different mathematical modeling approaches to studying ECM and cell-ECM interactions, ranging from continuum to discrete frameworks. Continuum models describe the densities of cells and ECM components with either ordinary differential equations (ODEs), typically used to describe total population size via a \rs{spatially-averaged} density, or partial differential equations (PDEs) that \rs{often} describe \rs{both} the spatial distribution and \rs{temporal evolution} of a population density. On the other hand, discrete frameworks, such as agent-based models (ABMs) represent biological components as \rs{separate, interacting agents.} ABMs can be further categorized as either lattice-based, where positions are restricted to a finite set of points, or lattice-free, where positions lie in a continuous range. In some cases, multiple modeling approaches are combined in a hybrid fashion to balance the computational complexity \rs{of frameworks with many interacting agents} with the accuracy of their outputs. \rs{Mechanical models are a subset of models focused on representing material properties such as stiffness and elasticity. Both continuum and discrete mechanical models are utilized to study the mechanical interactions between cells and the ECM.}}
    \label{fig:schematic-multiscale}
\end{figure}


\textbf{Continuum models.} 
\rs{Continuum models describe the density of cells and ECM constituents as smoothly-varying continuous functions (Figure \ref{fig:schematic-multiscale}), with models of cell movement and interactions with the ECM taking the form of either ordinary differential equations (ODEs) or partial differential equations (PDEs). ODE models are used to represent the evolution of a single quantity in space or time (cell density, for example), but are often limited in their utility as they cannot describe evolution in both space and time, as is often desirable for biological processes. However, in many contexts, a series of ODEs can be coupled to describe complex biological systems such as cell signaling pathways in cancer (\cite{itani2010ode}). Simple PDE models, on the other hand, represent cell density, $n$, as a scalar field, $n(x,t)$, where $x$ denotes position in one-dimensional space and $t$ represents time. The evolution of cell density has been well-studied via a class of PDE models called reaction-diffusion equations. Perhaps the most famous model for this is the Fisher-Kolmogorov-Pietrovskii-Piskunov (Fisher-KPP) model \citep{fisher_wave_1937, kolmogorov1937study}, whose numerical and analytical solutions have been studied extensively in the context of cell biology \citep{maini2004traveling, gerlee2016travelling}.
However, these models only explicitly consider changes in cell density with respect to time.
More sophisticated models for cell movement can be constructed by coupling cell and ECM densities to capture their chemo-mechanical interactions. 
Some continuum models are exactly solvable}, \citep{petrovskii2005exactly}, however, in more complex cases \rs{where exact solutions cannot be obtained}, there exists a suite of numerical techniques to approximate solutions at a relatively low computational cost. In general, \rs{ODE and PDE models} are amenable to mathematical analysis, for example using boundary layer techniques \citep{farrell2000robust}, or asymptotic methods \citep{keller1995asymptotic}, wherein the behavior of a system under certain limits \rs{(for example, long-time behavior)} may be studied. 

When conducting studies at the tissue scale, both the cells and individual ECM components are of a small size relative to the characteristic length scale of tissue, and are often densely packed. Thus, in many cases, a continuum framework for modeling spatially averaged behavior at the cell scale and its effects at the macroscopic scale of the tissue is often appropriate. 
An important consideration, however, is that continuum-based models often overlook smaller scale interactions. If small-scale behavior is also to be studied, \rs{it can be advantageous to consider model frameworks that capture behavior at higher spatial resolutions, such as discrete models}.


\textbf{Discrete agent-based models.} 
Agent-based models (ABMs), also referred to as individual-based models, establish independent agents that each interact and evolve according to a specific set of rules and predetermined behaviors \citep{metzcar2019review}.
\rs{At the cell scale, ABMs typically represent cells as agents, and in this context are sometimes referred to as cell-based models.} However, agents can also represent components of the ECM, and can be assigned characteristics such as size, shape, and polarity. ABMs can be broadly split into two categories: lattice-free and lattice-based models (Figure \ref{fig:schematic-multiscale}). Lattice-free models are often center-based, approximating the cell surface as a sphere (or a circular disc in 2D models). \rs{This approach does not confine the cells to a specific set of points in the domain, such as a lattice, and allows free movement of the cells in any direction.} Other popular lattice-free approaches include vertex models, where cells are modeled as polygons, and subcellular element models, where each cell is described as a collection of spheres or circles \citep{metzcar2019review}. Conversely, lattice-based models restrict cell position and movement to a lattice, where space is represented as a discrete set of points. \rs{In restricting position, lattice-based models are often more computationally efficient than their off-lattice counterparts. However, this simplicity can come at the expense of accuracy, particularly in processes where the position of agents is a major determinant of system behavior}. Commonly used approaches include cellular automata, where each cell is represented as the occupation of a lattice site, and Cellular Potts models (CPMs), where each cell is a collection of multiple lattice sites.

\rs{ABMs excel both at modeling stochastic environments and giving a fine-grained investigation of cell-cell mechanical interactions and dynamics of cell phenotypes (e.g. motile versus immotile cells).}
Due to their focus on individual cell behaviors, ABMs are generally better equipped to capture cell-level processes when compared with continuum models. 
Moreover, ABMs can incorporate subcellular processes which inform and drive cell behavior \citep{letort_physiboss_2019, ponce-de-leon_physiboss_2023, verstraete_agent-based_2023}. 
By representing cell-cell communication in models, agents can \rs{also} interact with one another to reproduce the emergence of population dynamics.
Disadvantages of using ABMs include that they are often computationally expensive and analytically intractable, particularly for large systems with many agents.


\textbf{Mechanical models.}
\rs{Mechanical models are a subset of models that place a focus on describing the material properties of biological structures. They can be applied both at the tissue or cell level, using either continuum or discrete approaches (Figure \ref{fig:schematic-multiscale}).}
\rs{We highlight this category due to its relevance for modeling ECM mechanobiology.}

\rs{At the heart of a mechanical model lie the constitutive equations, mathematical relations that} describe how a material responds to deformation (strain) caused by an applied force (stress) \citep{Chaudhuri2020}. 
The types of constitutive equations range in complexity from simple isotropic linear models \rs{(e.g. a linear spring)} to complex non-linear models that consider \rs{how anisotropy in the microstructure, for example ECM fiber alignment, influences the material response to forces applied from different directions}. 
\rs{Continuum mechanical models typically represent biological structures as geometrical objects onto which} continuous fields of stresses and strains are mapped \citep{guo2022modeling}, with the Finite Element Method (FEM) a popular choice for solving such stress-strain problems.

Generally, continuum mechanical models \rs{average tissue microstructure, making it challenging to represent highly diverse environments such as ECMs with spatially varying architecture, or tissue regions with different cell types. Inhomogeneous and anisotropic constitutive equations can to some extent address this limitation by incorporating spatially-varying and direction-dependent terms.} Further, continuum models generally assume that displacement within the modeled domain is negligibly small, an assumption that is violated when cellular forces substantially rearrange the ECM. 
In these cases, a more suitable approach is to break up the domain into smaller discrete regions. \rs{Discrete mechanical models typically represent cells or ECM as discrete objects, such as points distributed in space that can be mechanically linked to each other.}
As in continuum mechanical models, a constitutive equation is chosen to describe how each discrete object physically behaves upon force loading. 
\rs{ABM frameworks that include mechanical interactions between agents are thus also a type of discrete mechanical model.} 
Discrete mechanical ECM models are particularly suited to investigate emergent biophysical properties of matrix polymers \citep{broedersz2014modeling}, and long-range mechanical cues through the microenvironment \citep{alisafaei2021long}.
\rs{However, they come at the price of lower analytical tractability and increased computational cost.}


\textbf{Hybrid models.} 
Many of the models above are tailored to a biological problem with a specific characteristic length or time scale. However, there is often a disparity in such scales, for example, the size of cells is typically within micrometer range, whereas the thickness of collagen fibrils are usually of a nanometer-scale size \citep{Siadat2022}. 
Biological questions with such differences in scales motivates the use of hybrid models (Figure \ref{fig:schematic-multiscale}). 
Hybrid models employ a combination of different modeling techniques, coupling continuum and discrete approaches to better capture the dynamics and evolution of a system \citep{Stephanou2016}. 
A key advantage of hybrid models is that the rules governing the evolution of an individual cell can be easily decoupled from the underlying dynamics of continuous ECM constituent densities, resulting in a more comprehensive model of cell-cell and cell-ECM dynamics than purely discrete or continuum equivalents. 
However, this extra layer of complexity can also introduce a number of disadvantages, the most notable being computational expense.

\section{Cell Migration} \label{cellmig}
\begin{figure}
  \includegraphics[width=\textwidth]{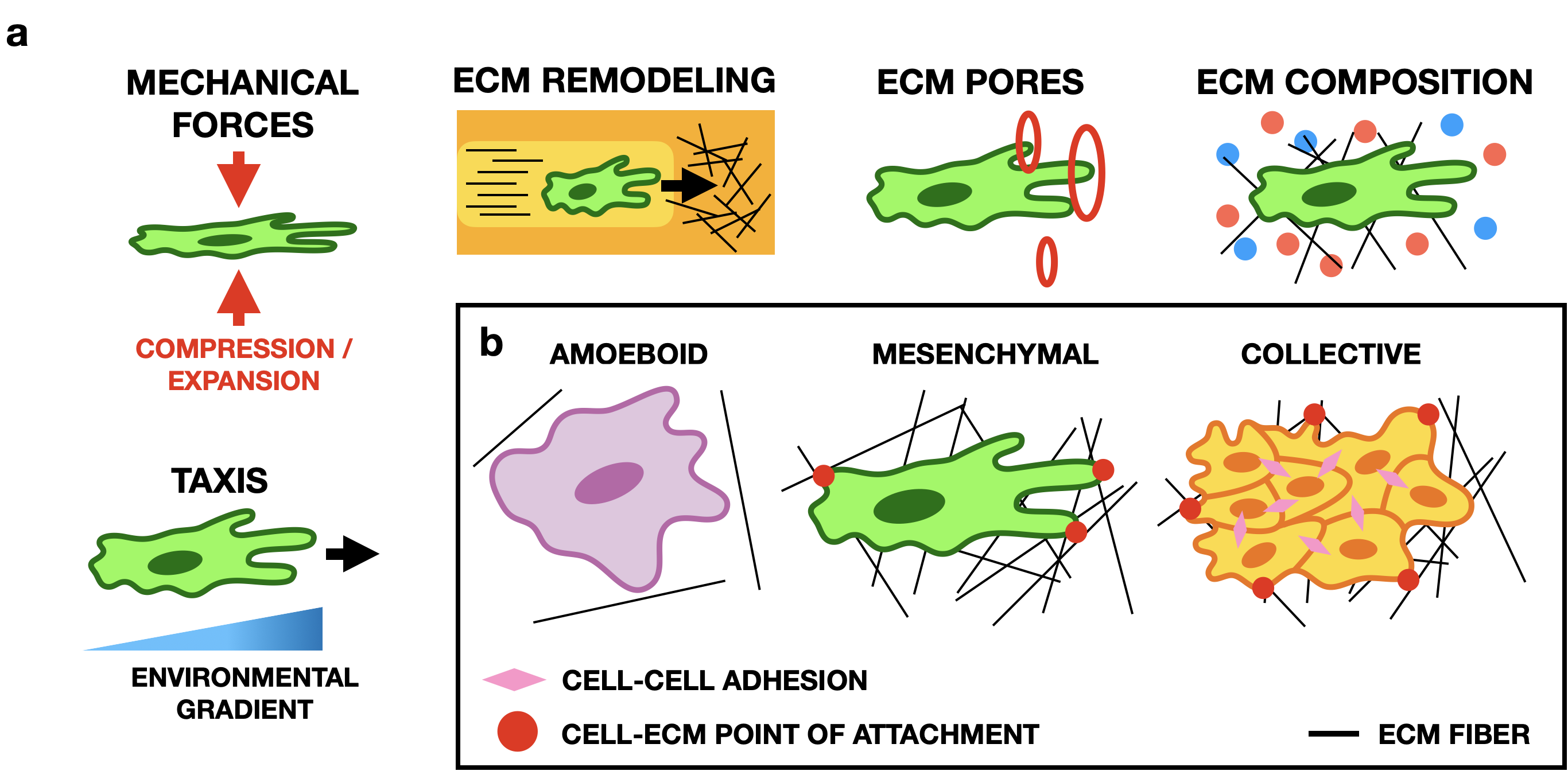}
\caption{\rs{ (a) Schematic illustrating typical cell-ECM interactions in migration. Cell morphology and migratory behavior are affected by many properties of the ECM, such as its composition, the forces it exerts on cells, the size of pores in the ECM, and gradients in the microenvironment of the cell (for example, the stiffness of the ECM or the concentration of potential points of focal adhesion). (b) Schematic showing common migratory cell phenotypes. Amoeboid cells form weak focal adhesions to the ECM, and move primarily by cortical actin flow and cell deformation. Conversely, mesenchymal cells form strong focal adhesions and pull on the ECM to move. In collective migration, cells mechanically adhere to one another and their microenvironment to facilitate cohesive movement as a group.}}
    \label{fig:cell_migration}
\end{figure}

The ECM influences the migratory properties of cells in many ways, including cell morphology, polarization, matrix deposition, and matrix degradation \citep{Saraswathibhatla_Indana_Chaudhuri_2023}.
Experimentally, cell morphology changes depending on the local geometry of the ECM and on the distribution of adhesion sites in the region through which a cell migrates \citep{wu_plasticity_2021}. 
Cells interact with the ECM through several types of actin-rich protrusions, such as filopodia and lamellipodia that form focal adhesions with the ECM to facilitate directed cell migration along environmental gradients (Figure \ref{fig:cell_migration}) \citep{caswell_actin-based_2018}. Here, we review mathematical and computational models considering the role of the ECM in cell migration during development, cancer cell migration, and wound healing.

\subsection{Mechanisms of cell migration}
\rs{The ECM impacts cell migration in a multitude of ways through both biochemical signaling and biophysical interactions \citep{Pally_2024}. Mathematical and computational models can unravel this complexity by selectively including only a subset of ECM components and their properties, and thereby explore the role of specific mechanisms in cell migration (Figure \ref{fig:cell_migration}).}

\textit{In vivo} tissues present a notable degree of spatial heterogeneity, primarily in the arrangement of ECM collagen fibers, \citep{yuan2016spatial, filipe2018charting}. \rs{The role of fiber orientation has been the focus of several models.} 
An early continuum model developed in the late 1990s describing cell migration through collagen-rich networks studies the relationship between fibroblast polarization and fiber orientation \citep{dallon1998mathematical}.
This work formed the basis of many subsequent models of cancer cell migration and wound healing that are discussed later in Sections~\ref{cancer}~and~\ref{woundHealing}.
Using a different approach to account for fiber orientation, \cite{Painter_2008} represents the ECM as a probability density function that includes a description of fiber alignment. 
\rs{Here, cell motion is modeled with run-and-tumble motility, in which the change in migration velocity and direction is given by a discrete random process, known as a `velocity-jump process' \citep{Othmer_1988}. In this model, the distribution of ECM fibers influences the rate and direction of each `jump'.} 
In amoeboid cell populations, their model shows that the structure of the ECM influences cell organization in the absence of any additional cues such as adhesion or chemotactic gradients. 
Furthermore, this model demonstrates that the realignment of the ECM fibers causes the formation of cell chains, a phenomenon also observed in mesenchymal-type cells such as neural crest cells \citep{Alhashem_2022}. \rs{Although this model permits ECM fiber reorientation, it does not model mechanical forces.}
%
To focus on the mechanics of fiber reorientation by cells, \cite{Schluter2012} develop an ABM. Their model predicts that stiffer ECM prevents the reorientation of fibers and that faster, persistent migration strategies emerge in environments with highly aligned ECM fibers. Furthermore, the model also suggests that fiber realignment by migrating cells promotes other nearby cells to trail behind, inducing leader-follower migration \citep{Qin2021}.

The ability of cells to change their shape to adapt to their surroundings is essential for migration and \rs{invasive capacity} \citep{van2018mechanoreciprocity}. Among ABMs, the CPM explicitly incorporates cell morphology\rs{, which \cite{Scianna2013_CPM_with_ecm2d3d} leverage to study the interplay between cell and nucleus deformability and ECM properties such as density, pore size, and fiber elasticity. They find that intermediate pore size permits cell passage while allowing optimal adhesion to ECM. ECM stiffness modulates this relationship: highly deformable soft ECM facilitates migration through small pores, while stiff ECM facilitates migration through large pores. Similarly, overly dense ECM prevents motility, while loose ECM reduces contact guidance. Follow-up work shows that a highly deformable nucleus promotes cell migration through constrictions in 2D and 3D \citep{scianna2021cellular}. }
To investigate the role of focal adhesions in cell migration, \cite{rens2020cell} develop a hybrid CPM--FEM framework. Here, cells can distinguish between soft and stiff ECM via the dynamic and force-dependent nature of focal adhesion growth, \rs{which is modeled by catch-slip bonds -- a type of interaction where the bond strengthens as it is pulled, and which is known to occur for integrin-ECM binding \citep{guo2022modeling}.}
\rs{As a result of self-reinforcing focal adhesion stabilization on stiffer substrates, cells spontaneously migrate toward stiffer matrix. Thus, this model proposes a mechanism for the emergence of durotaxis.}

\rs{Several works focused on detailed mechanical modeling of cell and ECM interactions in migration, for example using frameworks such as the subcellular element model \citep{Newman2007}. 
A highly detailed 3D ABM represents the cell surface as a collection of membrane elements that mechanically interface with the nucleus via elastic cytoskeletal strings \citep{He_2017}. This model predicts more persistent cell migration on concave surfaces than convex surfaces due to the different contact angle of adhesions.
Another model by \cite{kim2018computational} imposes durotaxis as an assumption, and studies the mechanism of stiffness sensing. In their 3D discrete mechanical model, a dynamically-shaped cell extends thin filopodia inside an ECM fiber network. Migration direction is biased towards the stiffest ECM sensed at the filopodial tips. This model reproduces experimental observations that cells are more likely to remain stationary when the distance between soft and stiff ECM increases, but steer towards stiffer ECM if they are within the reach of filopodia. This model was later extended to integrate intracellular mechanosensing \citep{kim2022computational}.
%
}

\rs{
In mechanosensing, intracellular cascades are triggered when cells adhere to the ECM through focal adhesions \citep{HANNA2013}. The ECM is considered a key regulator of many intracellular signaling pathways, with it being noted that varying ECM density can alter the activation potential of proteins in the Rho-family of small GTPases \citep{Bhad_2007}. 
Mathematical models have been developed to explore these complex signaling networks and uncover intracellular dynamics; we refer the reader to \citep{Hastings_2019} for a concise review of models of ECM-mediated signaling pathways.}

\rs{
One of the strengths of models is their ability to predict the emergence of complex behaviors from simple underlying mechanisms. 
Different models can lead to the same conclusion, such as the more abstract continuum model by \cite{Painter_2008} and the discrete model by \cite{Schluter2012} that accounts for mechanics. Both these models predict that mutual feedback between cell migration direction and ECM fiber orientation leads to the spontaneous emergence of cell chains and leader-follower behavior. This highlights how fiber reorientation is a critical mechanism, regardless of how it is achieved.
%
In other cases, more detailed models are necessary to recover certain features. Accounting explicitly for cell deformation allows models such as the CPM to study the influence of matrix microstructure such as pores. As a general trend recovered by such models, many ECM properties influence cell migration in a non-trivial manner, with an intermediate optimum \citep{Scianna2013_CPM_with_ecm2d3d, scianna2021cellular}. However, to predict durotaxis, a more accurate biomechanical model of focal adhesions is required \citep{rens2020cell}.
Highly detailed mechanical models aim to comprehensively describe the physics of the cytoskeleton, membrane deformation, and adhesion to the ECM \citep{He_2017, kim2018computational, kim2022computational}. These efforts, however, come at a price of a large amount of unknown parameters which limits model interpretability, and considerable computational cost of simulation.
Finally, from a molecular standpoint, models including mechanosensing can be developed through the integration of intracellular signaling pathways, such as those involving Rho GTPases \citep{Hastings_2019}. 
%
}

\subsection{Neural crest cell (NCC) migration}
The collective migration of neural crest cells (NCCs) underpins a range of developmental processes during embryogenesis \citep{mayor2013neural}. 
Defects in neural crest biology can result in a range of congenital diseases and birth defects termed neurocristopathies \citep{vega2018neurocristopathies}. 
Mathematical models are of particular interest in the context of NCC migration, often directing experiments towards important parameter ranges and identifying particular points of interest in the development of therapies that aim to prevent or counteract the effects of disrupted or misdirected migration. 
For a comprehensive review of mathematical models of NCC migration, we refer the reader to the review by \cite{Giniunaite2020}.
In this section, we briefly outline two recent hybrid models that explicitly consider the ECM in this context. 

One approach to modeling the migration of NCCs considers the influence of fibronectin network remodeling in NCC migration \citep{martinsonDynamicFibronectinAssembly2023}. 
Using PhysiCell, an open source physics-based cell simulator \citep{Ghaffarizadeh_Heiland_Friedman_Mumenthaler_Macklin_2018}, 
cells are modeled as freely moving 2D agents, whose velocities are functions of friction, cell-ECM interactions, and cell-cell repulsion. 
Simulations of this model predict that fibronectin deposition and remodeling by cells at the leading edge of collectives is an important mechanism in facilitating long-distance migration by preventing \rs{the collective arrest of cell motion, known as cell jamming \citep{Keister_2021}}.

A later \rs{developed} hybrid model for collective migration in the cranial neural crest describes the ECM as a scalar field that is degraded by migrating NCCs \citep{mclennan2022colec12}. 
The speed and direction of NCCs are influenced by the ECM in a tunneling mechanism that facilitates directed movement through regions of remodeled ECM. 
This model highlights the importance of confinement in cranial NCC migration, suggesting that factors such as Colec12 and Trail, which are expressed primarily in NCC-free zones adjacent to cell collectives, play a key role in confining NCCs during migration and maintaining coherence within collectives. 

\rs{In the context of NCC migration, these recent hybrid modeling developments reveal the importance of fibronectin and alignment of fibers in the ECM in guiding long-range migration.}


\subsection{Cancer cell migration and tumor development}
\label{cancer}

During cancer cell migration, changes in the composition, stiffness, and topography of the ECM can significantly influence cell behavior and migratory capacity \citep{najafi2019extracellular, park_topotaxis:_2018, eble2019extracellular}. 
Cancer cells remodel the ECM to facilitate migration \citep{mohan2020emerging}, for example by secreting enzymes such as MMPs \citep{castro2016cellular}. 
Additionally, changes in ECM structure can induce a more invasive phenotype \citep{leight2017extracellular}, further promoting migration.
In this section, we review models of cancer cell migration that explicitly consider the ECM during tumor progression. 

In mathematical studies of cancer cell migration, traveling waves are often used to describe the speed and shape of spatial processes in cancer progression \citep{gerlee2016travelling}. 
A simple continuum model that has been utilized extensively in the context of cell migration is the Fisher-KPP model \citep{fisher_wave_1937, kolmogorov1937study, maini2004traveling}. 
This seminal continuum model only explicitly considers changes in cell density with respect to space and time, and does not explicitly consider cellular interactions with the ECM. 
To more closely align theoretical predictions with \textit{in vivo} and \textit{in vitro} experiments, more sophisticated traveling wave models explicitly represent both cells and the ECM as distinct, smoothly-varying quantities that interact with one another. 
Newer models of this nature suggest that the speed of tumor invasion depends on the ECM density ahead of the invading front, and on the rate of ECM degradation by motile cancer cells \citep{el2021travelling, colson_travelling-wave_2021}. 
Similar results have also been found using a PDE model derived from the underlying individual cell-cell and cell-ECM interactions \citep{crossley2023travelling}, which, unlike the majority of continuum models, considers the volume-filling effects of both the cells and ECM during invasion. The resulting model describes non-linear cross-diffusion \citep{simpson2009multi}, and proliferation terms akin to those considered in a different model for melanoma invasion into human skin \citep{browning2019bayesian}. 
By considering asymptotic parameter regimes, this work justifies the use of simpler models such as the Fisher-KPP model to understand qualitative behaviors of model solutions, and employs boundary layer analysis to profile the traveling wave solutions for small ECM degradation rates. 

In addition to the ECM, there are several other extrinsic factors that also influence cell behavior during migration. A recent model describing acid-mediated tumor invasion \citep{strobl_mix_2020}, considers the role of both the ECM and stroma in impeding tumor cell movement and inhibiting growth. Tumors often consist of a heterogeneous population of cells \citep{sinha2020spatially}, so this model considers two tumor cell types that are able to antagonize either the ECM or stroma, and subsequently constructs a model of five coupled differential equations as an extension of Gatenby-Gawlinksi models of acid-mediated invasion \citep{gatenby2006acid, martin2010tumour}. The model predicts that heterogeneous tumors are more invasive than spatially separated tumor populations. Additionally, it concludes that the biological barrier of the stroma provides a stronger prevention of tumor growth than the physical ECM barrier.

The ECM not only influences cell migration, but also tumor morphology and its expansion. In this context, multi-scale moving boundary approaches are employed to understand behaviors across scales, ranging from the cell scale to the tissue scale. 
An initial framework introduced by \cite{Trucu_2013} is later extended to examine tumor morphology by considering the direct interaction of cancer cells with a two-phase ECM comprised of both a fibrous and non-fibrous component \citep{shuttleworth2019multiscale}. 
The inclusion of bi-directional cell-ECM interactions, including cell adhesion and dynamic fiber reorientation, enables an exploration of pattern formation, highlighting the characteristic fingering pattern observed in tumor modeling \citep{drasdo2012modeling}. 
By considering different compositions of ECM, such as varied initial fiber distributions and orientations, the authors show that tumors display more aggressive growth patterns when seeded in a heterogeneous ECM.

While continuum models are excellent at capturing population trends such as collective migration, ABMs are more suitable for lower cell densities, such as in single cell migration or for modeling the transition between collective and individual cell migration. 
\cite{Harjanto_2013} develop a 3D cellular automaton model with explicitly modeled ECM fibers to investigate four types of cell-matrix interactions: collagen-density dependent cell-mediated deposition, degradation, realignment, and displacement of collagen fibers. 
After parameter fitting to two prostate cancer cell lines, the model reveals an elevated probability of collagen degradation and motility even in denser ECM, which correlates to a more invasive tumor and matches experimental observations.
%

In an effort to make ABMs more accessible, computational packages for hybrid CPM such as CompuCell3D \citep{Swat_CC3D, izaguirre_c_2004} and ABM software such as PhysiCell \citep{Ghaffarizadeh_Heiland_Friedman_Mumenthaler_Macklin_2018} 
facilitate model development with minimal programming experience. Recent work in PhysiCell represents the ECM as a drag force to study how cancer cell migration and tumor cluster formation are affected by a 3D ECM architecture \citep{gonccalves2021extracellular}. 
This work also introduces a representation of the distribution of cell locomotive forces and intercellular adhesion and repulsion to capture the spread in cell velocities found \textit{in vitro}. This model predicts a dichotomy between cell migration and tumor size: if the former is hindered by increased ECM density, multicellular clusters increase in area, and vice-versa, such that even tumors with low levels of proliferation can be highly invasive.
A PhysiCell extension, PhysiBoSS, allows for the tracking of intracellular dynamics such as gene regulatory networks \citep{letort_physiboss_2019, ponce-de-leon_physiboss_2023}. \cite{ruscone_multiscale_2023} use PhysiBoSS to model cancer invasion through the ECM, which is represented as a scalar field in the base structure of the microenvironment. 
The ECM acts as a barrier for cells and cell-ECM interactions, such as repulsion and adhesion, along with intracellular regulation, are modeled using Boolean logic. This model successfully reproduces experimental results and demonstrates non-reversible \rs{epithelial-mesenchymal transitions (EMTs), a phenotypic switch from epithelial cells to mesenchymal cells, wherein cells lose cell-cell adhesion capabilities and gain both migratory and invasive properties \citep{Kalluri_2009}.}

\rs{Cancer cell migration has long been framed by the concept of EMT. 
Recent studies taking inspiration from the physics of active matter reframed tumor cell migration as state transitions, where a non-migratory tumor is likened to a solid or glass-like material, and a tumor exhibiting more migratory cells to a liquid or gas. 
In analogy to granular materials, this transition from immotile to motile is called the jamming-unjamming transition \citep{oswald2017jamming}. 
Interestingly, at least in some experimental systems, distinct biological mechanisms underlie EMT and jamming-unjamming transitions \citep{mitchel2020primary}.}
\rs{A key regulator of EMT is the cell-cell adhesion protein E-cadherin, whose loss of expression is associated with increased tumor invasiveness. 
To study how breast cancer cells switch from collective to single-cell migration, \cite{Ilina_2020} complement their experimental study with a discrete lattice-based cellular automata model based on a previously-developed ABM framework \citep{Deutsch_2021}. Key parameters in this model are cell-cell adhesion mediated by E-cadherin, and ECM density which confines cell movement. Increasing intercellular adhesion favors collective motion but also reduces cell migration speed, while increasing confinement by dense ECM reduces cell migration speed and forces cells into a solid-like jammed state. When both ECM density and intercellular adhesion are low, cells tend to move as individuals in a gas-like state. When ECM density is high and intercellular adhesion low, cells move collectively in loose flocks in a fluid-like state.
}

\rs{As in non-malignant cell populations, cancer cells utilize the ECM to facilitate migration. Models agree with experimental data that cancer cells adhere to dense ECM structures at a stronger rate than to soft ECM, where they secrete high levels of MMPs that degrade the surrounding matrix, which in turn frees up space for cell migration \citep{Harjanto_2013}. On the other hand, in a soft ECM, cancer cells are less likely to adhere to the matrix and therefore have a slower migration speed. 
Mathematical modeling has also been used to shown that heterogeneity in either the ECM or tumor cell population can increase tumor invasiveness \citep{shuttleworth2019multiscale, strobl_mix_2020}.
The vast array of mathematical models discussed here highlight the power of models to predict cellular responses to a range of mechanical ECM properties, and by understanding how tumors spread in surrounding tissue, we can develop therapies aiming to reduce the speed at which tumors grow and invade healthy tissues.
}


\subsubsection{The role of ECM remodeling enzymes}
Many mathematical models of cell-ECM interactions, including those discussed previously, assume that cancer cells themselves are responsible for degrading and remodeling the ECM.  
MDEs are generally localized: either on the cell membrane, at the tips of invadopodia \citep{weaver2006invadopodia, lu2011extracellular}, or with very small diffusivity \citep{werb1997ecm}.
Thus, the highly localized degradation of ECM by cells is a suitable approximation to employ, in particular because the degradation of MMPs occur on a much shorter time scale than degradation of ECM \citep{perumpanani1999extracellular, webb1999alterations}. 
However, some MMPs are freely diffusible, allowing degradation of the ECM without direct cell contact \citep{cabral-pacheco_roles_2020}. 
This behavior can be modeled by introducing a diffusible population of MDEs which are secreted by cells and degrade the ECM \citep{Anderson_2000}, and analyzed using partial integro-differential equations to describe the spatial-temporal dynamics of cancer invasion \citep{Chaplain2011}.

Many CPMs are hybridized with PDEs to model diffusible MDEs. An early example by \cite{szabo2012invasion} combines both diffusible and immobile ECM components to investigate how cell-ECM adhesion, cell motility, and ECM degradation impact invasive dynamics into an aligned fiber array. 
Whilst cell-ECM adhesion can lead to invasion as cells progressively spread to maximize their adhesive contact with the ECM, the addition of cell motility accelerates the invasive front. 
The effect of ECM degradation then depends on the relative strength of cell-ECM adhesion; weakly adherent cells preferentially migrate along tracts where matrix has been degraded, while strongly adherent cells prefer to migrate along ECM-rich paths.
\rs{Similarly, in their model of cell migration, \cite{Scianna2013_CPM_with_ecm2d3d} find that matrix degradation can enhance migration when the ECM is too dense or pore size too small. In contrast, ECM degradation is deleterious when the ECM is too sparse, as some ECM contact is necessary for cells to move.}
Using a similar model, \cite{Pal_manipulatingECM_controlLocalCancerInvasion} study how different ECM patterns affect migration. 
Varying adhesion parameters and the elasticity of fibers in the network is found to change the invasive potential of a cluster of cells. 
Moreover, randomly curved fibers decrease invasion while wave-like and parallel fibers increase cancer invasion speed and distance. 
Further work by \cite{Kumar2016_collectiveinvasionbyECMdensity_CPM} shows that increased MMP secretion and fiber alignment enhances cell migration, while increased ECM density inhibits the migration of cells. 
This implementation has since been extended to study how differences in cell size and deformability promote invasion during cancer cell migration \citep{Asadullah2021_heterogeneityinsizeanddeformability_CPM}. 
In a similar hybrid model, \cite{pally2019interplay} adapt the rules specifying the agent morphology such that the BM forms clusters, while collagen forms elongated structures. 
\cite{pramanik2021matrix} later extended this work to systematically investigate the impact of cell proliferation, and MMP diffusivity and inhibitor cooperativity, revealing five migratory phenotypes: non-invasive, dispersed individual invasion, multimodal invasion with both non-invasive and dispersed individual invasive cells, invasion as a non-adherent flock, and invasion as an adherent cluster.

In addition to MMPs, LOX is another important diffusible ECM-remodeling enzyme.
A continuum framework by \cite{Edalgo2018} couples a PDE for cancer cell movement to two PDEs governing the evolution of randomly oriented and crosslinked ECM fibers, both represented as distinct scalar fields. They include reaction-diffusion PDEs to capture ECM degradation, remodeling, and fiber crosslinking by MMPs and LOX. 
By distinguishing crosslinked and randomly oriented ECM fibers, the model reproduces the enhanced migration of cancer cells in the presence of LOX-induced crosslinking. 
This work highlights the importance of LOX in facilitating cancer cell migration and metastasis \citep{cox2013lox}, with implications for the developments of novel therapeutics combating the progression of cancer.  

The importance of LOX in remodeling the ECM to facilitate cancer cell metastasis has also been studied using CPMs and other seminal hybrid ABM-continuum frameworks \citep{Anderson_2000} that describe tumor growth within the microenvironment \citep{nguyen2019hybrid}. 
All ECM components, including collagen fibers and enzymes such as MMPs and LOX, are modeled using PDEs, with cancer cells modeled as discrete agents. 
Simulations show that the rate of invasion by cancer cells is higher in a uniformly distributed fiber concentration, rather than a random distribution, and that as fiber concentration increases, pore size increases and migration is more efficient, agreeing with experimental studies such as \cite{paul_cancer_2017}. 
As in prior work by \cite{Edalgo2018}, the study by \cite{nguyen2019hybrid} highlights the importance of LOX in migration and highlights LOX inhibition as a potential therapy to reduce cancer metastasis. 

\rs{Remodeling ECM enzymes, such as those that can degrade the matrix (MMPs) or those that help remodel the matrix (LOX), are crucial to tissue homeostasis. However, they are also exploited by cancer cells to help them invade the surrounding tissue. Through the up-regulated secretion of MMPs, cancer cells are able to degrade large amounts of the ECM which frees up space for proliferation and migration. Another important ECM enzyme is LOX, which plays a key role in cross-linking collagen with elastin fibers. Through mathematical modeling, it has been shown that migration of cancer cells is enhanced in the presence of LOX-mediated fiber cross-links, highlighting the importance of LOX during tumor invasion.}

\subsubsection{Tumor spheroids}
Continuum mechanics-based models are often used to study 3D tumor spheroids embedded in fibrous collagen gel.
Chemo-mechanical free energy of the ECM, cells and adhesion determines cell evolution in an ODE model by \cite{Ahmadzadeh_twowayfeedback}, where the ECM is modeled with radially aligned fibers using stress–strain relations for transversely isotropic materials. 
The work shows that below a predicted critical elastic modulus for the ECM, cells favor adhesion and remain within the tumor, whereas above the critical value, it is energetically favorable for cells to detach from the spheroid.
Similar energy-based approaches to modeling mechanical deformations, motor binding energy and mechano-chemical feedback are used to predict age-related differences in ECM structure that impact melanoma progression \citep{kaur2019remodeling}.

\rs{Experimental results have shown that various breast cancer cell lines demonstrate different modes of cell invasion. Confirming earlier findings from \cite{Ilina_2020} and using new experimental results in spheroids, \cite{Kang_NovelPhaseDiagram_TumorInvasion} present a hybrid model combining vertex and particle-based approaches for spheroid growth within an ECM. They propose a jamming phase diagram to describe how two key parameters, ECM confinement and cell motility, control the transition from a solid-like jammed state, to a liquid-like collective migration in loose strands, up to a more gas-like migration as single detached cells.}

A model by \cite{caiazzo2015multiscale} is used to simulate ECM fibers, and extended to model blood vessels and cells using a 3D ABM coupled to an FEM solver for oxygen evolution in time and space \citep{macnamara2020computational}.
The interactions between agents in this model are mainly mechanical, showing that tumor shape is driven by local structures. 
Cells grow in alignment with the fibers and around existing vasculature, and migration is led by interactions with the microenvironment, including ECM fibers, and as cells seek out regions of higher oxygen concentration.
An alternative approach applied to glioblastoma spheroid growth uses a mechanochemical model, where the ECM is modeled using a PDE that describes its diffusion towards the tumor spheroid core and its uptake by cells \citep{carrasco2023mechanobiological}.
An ODE describes nutrient-dependent cell proliferation and death, and further PDEs model the mechanical properties of cells within the spheroid. Through parameter analysis and validation against experimental data, the authors conclude that this mechanical model effectively approximates glioblastoma spheroid growth and shrinkage.

\rs{These results, similarly to other cell migration contexts, demonstrate that ECM elasticity and fiber alignment strongly influence tumor spheroid growth - both in speed and shape. Additionally, age-related structural and density changes in the ECM will impact modes of tumor progression across various cancers, where this observation can be successfully reproduced using experimental data.}

\subsection{Wound Healing}
\label{woundHealing}
Wound healing is essential for restoring tissue integrity and function after an injury and requires a cascade of coordinated events at both the cell and molecular levels.
Human skin, for example, is composed of two distinct layers - an outer layer, the epidermis, and an inner layer, the dermis \citep{mcgrath2004anatomy}. 
The epidermis is mainly composed of keratinocytes, while the dermis is predominantly composed of ECM, but also contains blood and lymphatic vessels, immune cells, and fibroblasts that locally deposit ECM constituents \citep{Pfisterer2021}. 
Skin wound healing proceeds in four major phases: haemostasis, inflammation, proliferation, and tissue remodeling \citep{Wilkinson2020}. Anomalies during any stage can cause improper tissue repair and complications, including significant tissue scarring \citep{guo_2010_wound}. 
Haemostasis, the first stage of wound healing, involves blood coagulation at the wound site promoted by platelets and fibrin fibers \citep{arnout2006haemostasis}. 
After coagulation, inflammatory cells clean the wound and generate chemokine gradients that attract keratinocytes, fibroblasts, and other cells towards the wound center \citep{koh2011inflammation}. 
In the proliferation phase, growth factors stimulate keratinocytes to proliferate and restore the epidermis \citep{landen2016transition}, while blood vessels are re-established by angiogenesis. The final phase consists of ECM remodeling \citep{GUERRA20181}. The ECM plays critical roles in coagulation, during migration of cells towards the wound center, and in the final remodeling stage.
In this section, we review modeling approaches that explicitly describe the interactions between cells and the ECM during wound healing. 
For in-depth reviews of broader mathematical models of wound healing and closure and wound healing angiogenesis, that do not necessarily consider ECM, we refer the reader to excellent reviews elsewhere \citep{Jorgensen2016, Flegg2015}.

Simple continuum mathematical frameworks represent cell density in wound healing as a scalar field. 
A seminal model considers the effect of insoluble, fibrillar ECM during wound-healing angiogenesis \citep{olsen1997mathematical}, where the ECM mediates the movement of endothelial cells towards the center of a wound \citep{lamalice2007endothelial}. 
Cells and the ECM interact with one another through haptotaxis \citep{ricoult2015substrate}, haptokinesis \citep{friedl2000biology}, ECM-mediated cell proliferation \citep{rabie2021matrix}, and ECM production and degradation \citep{winkler2020concepts}. 
Steady-state analysis and traveling wave solutions predicts that the ECM and cell densities evolve with approximately constant speed prior to reaching the center of the wound, with cell density increasing beyond baseline levels at the wound edge before settling to a pre-wounding equilibrium \citep{dale_speed_1994}. 


ECM orientation and alignment are also of importance during the healing process and can influence the invasive capacity of cells during migration towards the centre of a wound \citep{lin2020mechanical, ray2021aligned}. Furthermore, ECM modulates the direction of cell migration through a process known as contact guidance \citep{dunn1976new}. 
Thus, an important extension to traditional continuum models is a representation of ECM fiber orientations, pioneered by \cite{dallon1998mathematical}. 
Models of this sort have been used to study fiber alignment as a dynamic and reversible process - first in two fixed orthogonal directions \citep{olsen1998simple}, and then for a continuous range of directions \citep{olsen1999mathematical}, in the context of angiogenesis.

To model the final remodeling stage of skin wound healing, \cite{dallonMathematicalModellingExtracellular1999} develop a hybrid framework with cells as discrete agents and collagen fibers as a continuous bi-directional vector field describing fiber density and directionality.
In the model, ECM orientation biases the direction of cell migration and migrating cells also reorient ECM. This reciprocal interaction is sufficient to generate fiber alignment in model simulations, with the precise nature of patterns depending on cell speed, the degree of cell polarization, and the initial fiber structure.  
\cite{cumming2010mathematical} extend this work by including both collagen fibers and fibrins, both modeled as continuous tensorial fields in 2D, and model fibroblasts as discrete circular discs, whose behaviors are determined by chemotaxis, volume exclusion, and contact guidance. This work successfully reproduces ECM remodeling dynamics observed both with and without scarring. 
Recent experimental work highlights the importance of viscoelastic behaviors in cell and tissue function \citep{Eroles2023, Huerta-Lopez2023}, which is subsequently incorporated into corresponding hybrid models \citep{Pensalfini2023}. 
To investigate dermal wound healing, \cite{Guo2022} couple a microscale stochastic cell adhesion model with a macroscale continuum mechanics ECM modeled as a fiber-reinforced material with hyper-viscoelastic constitutive behavior to represent fibroblast contraction.
This model qualitatively captures the cutaneous wound healing process, with initial active contractions, increased fibroblast population at the wounded site, and consequential stress distribution changes.

To model epidermal-dermal interactions during wound healing, \cite{wang2019multiscale} model the epidermis and dermis as separate compartments. 
Keratinocytes in the epidermis compartment are modeled as individuals, whereas fibroblast and immune cells in the dermis compartment are modeled as continuous fields using Keller-Segel reaction-diffusion-advection PDEs \citep{arumugam2021keller}. 
Diffusible signals across the BM, such as signaling molecules for ECM deposition and degradation are also modeled explicitly as continuous quantities. 
The model predicts that dermal wounds heal by forming scar tissue with different ECM and fibroblast compositions than healthy tissue, in agreement with experimental studies \citep{pastar2014epithelialization}. Furthermore, the model predicts wound depth to be a critical determinant in scarring. In particular, shallow but wide wounds are predicted to repair with smaller scars than narrow but deep wounds \citep{hinshaw1965histology, marshall2018cutaneous}. 
The benefit of this compartmentalized framework is that it is easily extensible without affecting existing model components, enabling a model representation of sharp boundaries within the skin such as the follicle/dermal boundary.  

The ECM also impacts regeneration in tissues other than skin. In a model of tendon healing, \cite{dudziuk2019simple} employ a continuous integro-PDE model to study the orientation of collagen fibers during scar formation. 
This work highlights the role of the initial fiber orientations in determining the final structure of the collagen network. 
Later work using a similar mathematical framework \citep{carrillo2021mathematical}, adds local alignment interactions between collagen fibers and simplifies this model under appropriate limits, allowing for a parameterization based on experimental data.

\rs{A large fraction of mathematical models currently investigating wound healing focus on the role of ECM fibers, in particular, fiber alignment and orientation. Through different mathematical approaches, these models agree that the ECM structure and configuration of the fibers facilitate cell migration and guide the cells towards the wounded site, contributing to the success of wound healing.}

\section{Tissue Structure and Morphology} \label{tissue-morph}

One of the major functions of the ECM is to mechanically support and maintain tissue shape both during tissue formation and in homeostasis \citep{walma2020extracellular}. The ECM provides not only mechanical support, but also regulatory input that can determine cell fates. For example, stem cells in mammalian skin need contact with the BM to retain their proliferative potential \citep{rousselle2022basement, shen2023rete}. Here, we review modeling approaches that focus on the ECM's impact on tissue structure (Sections \ref{cell_ecm_mech} and \ref{integrity}) and its effects in patterning and morphogenesis (Section \ref{morphogenesis}).

\subsection{Cell-ECM mechanics and ECM structure}
\label{cell_ecm_mech}

A fascinating observation is the transmission of long-ranged cellular forces via the ECM. Experimental work in the 1980s showed that fibroblasts exert sufficient force to densify and align collagen fibers over distances much larger than the cells themselves, and that these aligned ECM tracts encourage cell migration \citep{stopak1982connective}. Cellular forces lead to the formation of so-called intercellular ECM bridges, which are visible bands of dense and aligned ECM between distant cells \citep{panchenko2022does}. 
These experimental observations inspired the development of Turing-type mechanochemical models. Turing models, first proposed by \cite{turing1952chemical}, are a class of pattern formation models that rely on the reaction and diffusion of chemicals. A well-known example consists of two chemical species, a short-ranged activator and a long-ranged inhibitor \citep{kondo2010reaction}. Mechanochemical Turing models, pioneered by \cite{oster1983mechanical}, substitute one or more chemical species by mechanical constitutive equations coupling cell and ECM fields. These models explicitly describe the ECM as a continuum field with variable concentration and displacement via diffusive, advective, and convective flows. Like its chemical cousin, the mechanochemical Turing model predicts the formation of spatial patterns of spots or stripes denoting regions of high cell and ECM density. 

In the seminal model by \cite{oster1983mechanical}, fibroblast contraction takes the role of both activator and inhibitor of the typical biochemical Turing model. On the one hand, contraction pulls in both nearby collagen and nearby cells; the local increase in cells and collagen leads to greater local contraction in a positive feedback loop, thus acting as a short-ranged activator. On the other hand, this leads to a decrease of both cell and collagen density at a longer range, thus acting as a long-ranged inhibitor. 
An extensive body of work has built on this, for instance, to include more complex cellular behavior such as cell division and haptotaxis \citep{murray1988mechanochemical, holmes2000mathematical, tranqui2000mechanical}. Models of this type are widely applied to \textit{in vitro} endothelial cell sprouting on ECM substrates \citep{manoussaki2003mechanochemical, murray2003mechanochemical, namy2004critical, tosin2006mechanics}, patterning of epidermal placodes and dermal papillae \citep{murray1988mechanochemical}, patterning during limb morphogenesis \citep{murray1988mechanochemical, oster1985model}, wound healing \citep{olsen1995mechanochemical, olsen1998spatially, tranquillo1992continuum}, and tumor metastasis \citep{tracqui1995passive}. Variations of the model have been developed to include other types of ECM mechanics such as osmotic swelling and de-swelling due to the secretion or degradation of hyaluronic acid \citep{oster1985model}. 
Several studies also analyze how the choice of constitutive equation for the cell forces and matrix displacement impacts the patterns predicted by the model \citep{byrne1996importance, villa2021mechanical}. 
Patterns consistent with mechanochemical Turing models are observed experimentally and are suggested to lay down the pre-pattern for skin follicle morphogenesis \citep{palmquist2022reciprocal}. 

Another widely-used approach to understand how cell forces spread through the ECM and ultimately form intercellular ECM bridges is based on mechanical models \citep{wang2020continuum}. 
Many models of this type consider a large (sometimes infinite) ECM gel seeded with a relatively small number of contractile cells, often modeled as shrinking circles or ellipses. Multiscale models combining a microscopic discrete mechanical fiber model with a macroscopic continuum ECM biogel model have also been used \citep{aghvami2013multiscale}. 
This body of work predicts various contributing mechanisms to intercellular ECM bridge formation, including cell elongation \citep{abhilash2014remodeling, han2018cell}, ECM fiber buckling \citep{notbohm2015microbuckling, han2018cell}, focal adhesion mechanics \citep{cao2017multiscale}, non-linear ECM fiber mechanics \citep{goren2020elastic, grekas2021cells, sopher2023intercellular}, fiber re-orientation \citep{goren2020elastic, panchenko2022does}, force-dependent crosslink breakage and plastic ECM deformation \citep{kim2017stress, ban2018mechanisms, ban2019strong, malandrino2019dynamic}. Clearly, many different mechanisms can qualitatively explain long-ranged force transmission. However, the precise ECM network architecture can quantitatively affect the range and heterogeneity of propagation of cellular forces through ECM \citep{humphries2017mechanical}. Thus, a close matching of simulated ECM to quantitative experimental data is paramount.


Although the aforementioned models successfully capture realistic ECM mechanics, they tend to oversimplify cell shape. To address this shortcoming, some approaches incorporate deformable cells coupled to discrete mechanical models of ECM networks.
%
Perhaps the earliest such model was developed by \cite{reinhardt2014agent}, in which cells are made up of multiple membrane elements that interface with fibers from a 2D ECM network model. In follow-up work, \cite{reinhardt2018agent} calibrate their model against experimental measurements of collagen biophysical properties. Both models successfully replicate the formation of intercellular ECM bridges. 
A similar approach is used by \cite{slater2021transient} and combined with a more sophisticated ECM network model that includes force-dependent crosslinker unbinding.
\cite{eichinger2021computational} hybridized an ABM with an ECM fiber network that is fitted to confocal microscopy data of collagen gels. In follow up work, \cite{paukner2023key} use this model to investigate the minimal criteria needed to obtain durotaxis. \rs{They identify two sufficient conditions for durotaxis. Firstly, the cell must adhere to ECM via catch-slip bonds. Secondly, the cell must continuously pull on their adhesions to the ECM via actomyosin contraction of its internal cytoskeletal network. 
Interestingly, both of these factors are included in a different model of a CPM cell on a continuum ECM substrate (discussed in Section \ref{cellmig} in the context of cell migration), which also predicts emerging durotaxis \citep{rens2020cell}. The fact that two models with such different implementations come to the same conclusion supports the notion that indeed, the biomechanics of integrin bonds and cytoskeletal activity are key to durotaxis. }
%
To allow even more cell shape flexibility, \cite{tsingos2023hybrid} hybridized a CPM with a discrete mechanical ECM model that captures long-ranged strain and force transmission through the ECM network. \rs{They highlight a transition from viscoplastic to viscoelastic ECM mechanical properties as the density of fiber crosslinking is increased.}

\rs{
Intercellular ECM bridges have been proposed to be involved in long-ranged sensing and potentially communication between distant cells \citep{Nahum2023, panchenko2022does}. By considering a range of mathematical approaches, the models highlight the key role of cell-derived forces and fiber realignment in the formation of intercellular ECM bridges. Furthermore, many different ECM mechanical properties are permissive to long-ranged force transmission, suggesting this mode of mechanosensing is robust to variations in ECM composition, and could be co-opted to regulate cell function in different tissue contexts. Indeed, long-ranged mechanical interactions mediated by ECM have been proposed to guide collective cell migration, orchestrate wound healing, and allow cells to coordinate during morphogenesis \citep{sapir2017talking}. The pattern-formation potential of relatively simple force-based interactions between cells and ECM are beautifully demonstrated by the Turing-type models. 
As continuum models, Turing-type models provide an excellent tool at tissue length-scales, while mechanical ECM models tend to focus on smaller cellular scales, but with few exceptions have not incorporated detailed cellular behavior models typically seen in ABM. Recent advances in coupling mechanical network models of the ECM with ABM frameworks such as CPM will enable to investigate how long-ranged sensing and communication impact on migration and pattern formation at the cellular level.
}

\subsection{Tissue integrity}
\label{integrity}

The role of ECM in maintaining tissue integrity has long been appreciated in biomedical engineering, where continuum mechanical modeling of biological tissues including ECM is applied to predict the effects of injury in joint cartilages, the cardiovascular system, and bones, as well as to design synthetic biomaterials for prosthetics, such as arterial stents and heart valve replacements \citep{khaniki2022hyperelastic}. The strong focus on biomedical engineering has led to the development of many sophisticated material models, for which the reader is referred to excellent reviews elsewhere \citep{guo2022modeling, holzapfel2019fibre}.

In many ABMs, ECM surrounding the tissue of interest is included by creating a new type of agent. Frequently, these ECM agents are made immobile and act as a physical boundary as well as attachment points for the freely-moving cells. Although clearly a simplification, this approach permits an investigation of how ECM structure affects cell behavior. For example, \cite{buske2011comprehensive} models the BM lining the mammalian intestinal epithelium by discrete points with pre-specified positions in the shape of a typical intestinal crypt. A key assumption of this model is that local tissue curvature affects cell proliferation via Wnt-signaling \citep{huelsken2002wnt}, enabling self-regulation of the proliferative compartment size.
%
Similarly, in a model of mammalian skin epidermis by \cite{sutterlin20173d}, cells are simulated using ellipsoids and the BM either as a flat or curved static surface. In this model, feedback regulation between surface skin layers and deep skin layers in contact with the BM ensures that the epidermis achieves a minimum homeostatic depth regardless of BM curvature, and showed that cell proliferation adjusts to fill deeper ECM clefts.
%
A static reticular network -- a structure made of close association of fibroblasts with ECM -- is used in a CPM of T~cell motility in the lymph nodes \citep{beltman2007lymph}. This model suggests that many features of T~cell motility such as stop-and-go motion can be explained by cells colliding with the maze-like reticular network.

Discrete mechanical models of laminin structure have been used to explore how BM integrity affects adjacent tissues, using for example a 2D honeycomb lattice \citep{reuten2021basement}. 
By removing nodes in this lattice, ECM softening by Netrin-4, a BM component associated with reduced metastases and better prognoses in cancer patients, is modeled. 
A similar laminin polygonal model is constructed to explain how laminin ECM aids in epithelial sheet migration during optic cup morphogenesis in zebrafish embryos \citep{soans2022collective}. \rs{The model proposes that weakening of the laminin matrix creates holes in the BM that prevent collective migration.}

\rs{The ECM plays a key role in maintaining the integrity of surrounding tissues. The models reviewed in this section simplify the ECM as a static barrier. Despite this strong simplification, such models permit the study of how the mere presence of a steric obstacle can regulate cell motility and proliferation dynamics. 
}

\subsection{Morphogenesis}
\label{morphogenesis}

Morphogenesis refers to the acquisition of tissue shape during development, which may involve formation of patterns of cell clumps, but also cell migration, cell adhesion, cell shape change, and differentiation -- all processes that can be regulated by the ECM \citep{walma2020extracellular}. In this section, we first review models of cell and ECM pattern formation before focusing on models considering the influence of ECM on the generation of tissue shape.

\subsubsection{Cell and ECM patterning}
Besides the aforementioned mechanochemical Turing models, other ECM-mediated pattern-forming mechanisms have been studied with computational models. 
%
A general study of pattern-forming potential uses a cellular automaton model in which each lattice site contains either a cell, ECM particle, or fluid medium \citep{grant2006simulating}. Phenomenological rules are used to explore how cells self-organize into different structures, including cysts surrounding a fluid-filled lumen, cysts surrounding an ECM lumen, or boundaries separating a fluid phase from an ECM interior like an epithelium.
In a more involved approach, \cite{checa2015emergence} combine a cellular automaton model with a continuum mechanics-based description of the ECM to study how cell behavior such as stiffness-mediated realignment coupled to ECM mechanics leads to the formation of patterns of aligned cells and fibers.

To study how patterned ECM emerges from fibroblast migration, \cite{Wershof2019_feedbackmartix} use a Vicsek model (a type of ABM) coupled to an ECM model on a lattice. Three mechanisms contribute to fibroblast motility: migration noise, cell-cell guidance, and cell-matrix guidance. Cells affect the underlying ECM via fiber deposition, realignment, and degradation. In the absence of cell-matrix guidance, the ECM is either completely disorganized (high migration noise, low cell-cell guidance), or organizes into long coherent strands (low migration noise, high cell-cell guidance). Cell-matrix guidance leads to the generation of wavy fibers and swirls reminiscent of patterns in various tissue types observed by the authors.

To explain the formation of cell clumps in their experimental \textit{in vitro} work on mesenchymal cell condensation, \cite{zeng2004non} develop a CPM where cells secrete fibronectin on the lattice sites they occupy, forming insoluble non-diffusing deposits. Cells perform haptotaxis toward fibronectin, and cell-fibronectin adhesion leads to the upregulation of cell-cell adhesion proteins. Rapidly and randomly moving cells progressively clump and slow down due to the self-reinforcing loop of fibronectin secretion and increased cell-cell adhesion. 
This model correctly recapitulates that cell seeding density is correlated with the transition from spots to stripes, and the fact that (unlike Turing patterns) the clump patterns are highly irregular.

White adipose tissue is organized into lobules separated by collagen-rich septa, which \cite{peurichard2017simple} sought to model with an ABM representing adipocytes as spheres and ECM fibers as lines.
They recapitulate the formation of lobules and septa by simple mechanical interactions; as islands of adipocytes expand by cell proliferation and growth, they compress collagen fibers into septa until the tissue is densely packed to an extent that no new cell growth can occur.

During development, the mammalian skin BM changes from an initially flat surface to an undulated one \citep{shen2023rete}. In this context, the interaction between cell proliferation and the ECM is also the focus of several models where the BM is represented as a collection of small spheres \citep{kobayashi2018interplay, ohno2021computational}. The authors argue that mechanical buckling alone cannot explain undulation, as buckling neither gives a direction of undulation nor does it lead to irregular undulations, as observed in the skin. In contrast, their model shows that cell division dynamics could bias the ECM curvature due to cells pinching off the ECM when detaching. In addition, the positioning of stem cells along the tips of the undulated BM emerges spontaneously due to the combined action of cell-ECM adhesion and cell division.

\rs{
Realignment of the ECM by migrating cells leads to the formation of patterns, an observation that has also been made in other models of cell migration \citep{Painter_2008}. In particular, cell-ECM guidance causes the organization of migrating cells into long coherent, directed strands. Cell clumping, on the other hand, is a result of a feedback loop between cell secretion of fibronectin and increased cell-cell adhesion. 
Different patterns can emerge by steric interactions due to cell proliferation. As shown by \cite{peurichard2017simple}, cell growth can compress strands of initially loose ECM to form thick septa. In the skin, cell proliferation may underlie the formation of undulations in the BM \citep{kobayashi2018interplay, ohno2021computational}.
}

\subsubsection{Blood vessel morphogenesis}

\begin{figure}
  \includegraphics[width=\textwidth]{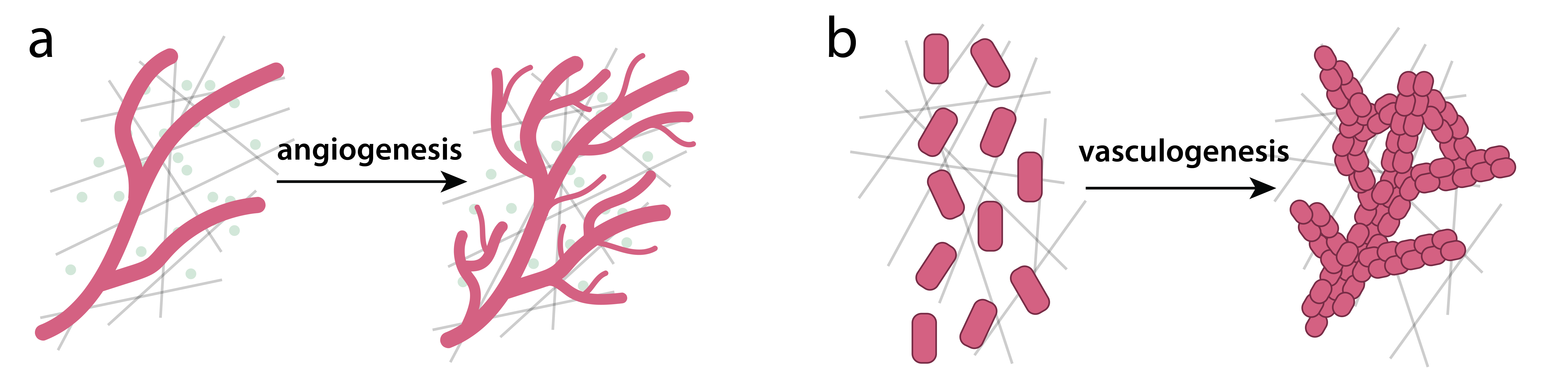}
\caption{\rs{Schematic depiction of angiogenesis and vasculogenesis. (a) Angiogenesis is a process in which new blood vessels form from a pre-existing network, whilst (b) vasculogenesis is the process by which randomly scattered cells spontaneously form vessel networks. During both of these processes, the underlying ECM plays an important role.}}
    \label{fig:genesis}
\end{figure}

\textbf{Models of angiogenesis}.
Angiogenesis refers to the formation of new blood vessels from pre-existing ones \citep{folkman2006angiogenesis} (\rs{Figure \ref{fig:genesis}a}). During angiogenesis, hypoxic tissues, such as tumors, secrete vascular endothelial growth factor (VEGF), which diffuses around cells and acts as a chemotactic cue for endothelial cells. Endothelial cells sprout from pre-existing blood vessels and migrate towards the source of VEGF. 
The ECM plays central role in angiogenesis and vascular remodeling, which has led to extensive theoretical work explicitly including the ECM. Here, we highlight some of these modeling works, and refer the reader to more in-depth reviews elsewhere \citep{heck2015computational, abdalrahman2022role, Crawshaw2023}.

The role of different cues in angiogenesis is a long-standing question. In this context, \cite{anderson1998continuous} study the balance between chemotaxis and haptotaxis. Their continuum model predicts that, when chemotaxis prevails, cells quickly reach their target tissue. However when haptotaxis dominates, the sprout moves slower and can fail to reach its target tissue. As the continuum approach cannot track individual cells, and thus precludes analysis of sprout morphology, the authors also derive a discrete cellular automaton model. In this model, migration via chemotaxis is predicted to favor straight sprouts, whilst haptotaxis favors the formation of sprouts looping onto themselves to form anastomoses. 
In contrast to \cite{anderson1998continuous}, \cite{levine2001mathematical} argue, based on experimental evidence, for \textquotesingle{}negative\textquotesingle{} haptotaxis, that is, cell movement from high to low substrate concentrations. In their continuum model, endothelial cells secrete MMPs in response to VEGF, locally degrading fibronectin. Thus, the cells at the migrating front carve a path for following cells, promoting further migration despite the absence of chemotaxis. 

The seminal ABM of \cite{anderson1998continuous} inspired many other ABMs of angiogenesis, including CPMs. 
\cite{bauer2007cell} introduce a model of interstitial ECM as line segments in the CPM lattice, and model how cells crawl on top of, adhere to, and degrade ECM. The ECM is a static obstacle around which cells must move, a task well-suited to be modeled with the CPM.
Here and in a follow-up model, \citep{bauer2009topography}, an optimum ECM fiber density is required for successful angiogenesis. In sparse ECM, degradation reduces sprout extension speed, while in dense ECM, degradation enhances sprout extension speed. However, beyond a threshold ECM density, sprout migration slows down considerably.
%
A similar ECM density optimum is also found in a hybrid model by \cite{milde2008hybrid}. They represent the ECM as a tensor with density and directionality in 3D, while the cells at the sprout tip are modeled as agents. 
The model predicts that low ECM densities lead to slowly advancing but unbranched vessels, intermediate ECM densities cause the vessels to advance quickly but form many branches, while at very high ECM densities, vessels advance slowly and branch frequently. 
In line with the model by \cite{anderson1998continuous}, increased fibronectin-directed haptotaxis leads to increased vessel tortuosity, as haptotaxis acts as a local autocrine directional cue.

\cite{daub2013cell} investigate the roles of MMPs in the balance of chemotaxis, haptotaxis and haptokinesis (faster cell speed toward denser ECM). The model combines CPM with a mechanical continuum ECM model, and predicts that strong chemotaxis favors long unbranched sprouts, while stronger haptotaxis or -kinesis favors short branched sprouts. These predictions are similar to those by \cite{anderson1998continuous}. 
\cite{boas2013computational} also study sprout morphology, but include both MMPs and soluble MMP inhibitors. At low secretion of both factors or high levels of MMP secretion, sprouts showed tubular morphology; at high levels of MMP activator secretion, sprouts developed into cyst-like structures; and high secretion of both MMP and their activators led to collective movement of the entire endothelial cell population without sprouting. In follow-up work the authors explored and experimentally validated the positive feedback loop between ECM degradation, and release of bound growth factors that further stimulate MMP production \citep{boas2018local}.

Several models focus on mechanical cues in angiogenesis. 
\cite{edgar2013computational} represent local ECM orientation as a vector field that biases the growth direction of blood vessels. They later expanded this model to simulate how macroscopic gel deformations affected local ECM fiber orientation and vessel sprouting \citep{edgar2015coupled}.
\cite{stephanou2015hybrid} model individual endothelial cells superimposed on a continuum chemo-mechanical model of the ECM. They include a strain-biased diffusion tensor, allowing matrix deformation to feed back directly on cell migration direction. They identify an intermediate parameter range for cell traction forces and ECM stiffness that enable angiogenesis.

\rs{Recent models have extended previous work by introducing Delta-Notch signaling pathways to model phenotype acquisition of sprout tip and stalk cells.}
\rs{Extending a model by \cite{vanOers2014mechanical} where a CPM is hybridized with a continuum mechanics model of the ECM, \cite{vega2020notch} introduce Delta-Notch-Jagged signaling. The model predicts that higher Delta reduces the number of sprouts and thickens them, while higher Jagged leads to more branching as it promotes lateral sprout induction. ECM mechanics play into this dynamic by promoting cell extension and lateral branch anastomosis. In follow-up work, the authors combine their angiogenesis model with a model of retinal cells and BM to simulate age-related macular degeneration \citep{vega2021anomalous}. They predict that impaired intercellular adhesion, excess VEGF and Jagged expression contribute to the disease.}
\rs{More recently, \cite{stepanova2021multiscale} develop a 2D hybrid, multiscale model for angiogenesis with phenotype-dependent cell-ECM interactions. 
At the subcellular level, a stochastic model of VEGF-Delta-Notch signaling determines two phenotypes: Delta-high tip cells and Notch-high stalk cells.
The phenotypes differ at the cellular scale, as tip cells are more motile, degrade interstitial ECM, and deposit BM. Interstitial ECM prevents cell migration, while the deposited BM prevents sprout branching.
Finally, the tissue scale employs ODEs to track changes in ECM component concentrations, and ECM fiber orientation. As in many other models, cell migration direction and ECM fiber alignment mutually influence each other.
In contrast to other models, here both branching and chemotactic sensitivity emerge dynamically. Further, this model identifies interstitial ECM degradation and BM deposition as crucial processes in determining number and length of vessel branches.
} 

A few common characteristics emerge in these studies. First, chemotaxis and haptotaxis are often predicted to have opposing effects on sprout migration velocity and morphology. Second, mechanical features -- ECM density, stiffness, cell forces -- display a non-monotonic relationship with cell migration, being optimal at an intermediate value. \rs{Finally, newer models have begun extending the scales to include biochemical signalling.} Nevertheless, due to the different mechanisms investigated in each study, comparing the validity of these models remains challenging. Quantitative comparisons of models to experimental data will be invaluable to identify the leading mechanisms of angiogenesis. 


\textbf{Models of vasculogenesis.}
An alternative way of blood vessel formation is vasculogenesis, wherein blood vessels are created \textit{de novo} from scattered endothelial cells (\rs{Figure \ref{fig:genesis}b}). Experimental vasculogenesis in the tube formation assay begins with isolated cells plated on ECM. The cells then coalesce into strands, forming a loose cell network \citep{arnaoutova2009endothelial}.
As mentioned earlier, mechanochemical Turing models have been applied to vasculogenesis \citep{manoussaki2003mechanochemical, murray2003mechanochemical, namy2004critical, tosin2006mechanics}. In general, these models predict the formation of stable patterns of cell strands, which correspond to positions where cell forces densified ECM.

To model the formation of vascular networks in embryos, \cite{kohn2011early} propose and experimentally validate a model where VEGF, secreted by adjacent tissue, can be immobilized through binding to the ECM \citep{kohn2013dynamics}. Thus, in contrast to angiogenesis, secreted VEGF does not form a gradient, but is uniformly present. In their model, randomly distributed CPM cells deposit ECM on the lattice sites they occupy. Because deposited ECM binds VEGF, sequestered VEGF locally accumulates. These local bound VEGF gradients induce cell chemotaxis, effectively encouraging cells to stretch and move towards locations they previously visited. 

Using a CPM hybridized with a continuum mechanics model of ECM, \cite{vanOers2014mechanical} model cell-generated ECM strains, while the reciprocal coupling of the ECM to cells occurs indirectly by biasing cell motility towards regions of higher ECM strain. Follow-up work by \cite{ramos2018capillary} shows that cells elongate and align along tracts of high strain, demonstrating that patterning requires an intermediate level of ECM stiffness
More recently, \cite{carrasco2023silico} study the effects of the ECM on vasculogenesis using a 3D ABM. 
They predict a strong relationship between ECM stiffness and vascular network proliferation, and in particular, that as the viscoelasticity of the ECM is increased, the vascular network proliferates less, such that network growth occurs primarily in regions of stiffer ECM. These predictions are partially validated through comparisons with existing experimental studies of vasculogenesis. 
\cite{noerr2023optimal} use an ABM in which cells align along directions of strongest ECM strain through the exertion of dipole forces. Three parameters are found to affect the ability to form a vascular network in the model: cell density, cell contractility, and substrate stiffness. Substrates that enable long-ranged force transmission are found to be most suited to lead to vascular network formation even at low cell densities.

\rs{The reviewed studies predict that the success of vasculogenesis strongly depends on the ECM. Although playing a different role during angiogenesis, VEGF facilitates vasculogenesis through local accumulation by binding to ECM, creating pockets of high VEGF concentration to which cells preferentially migrate towards. 
Another common theme among models is the role of ECM mechanics, as vasculature development requires some level of ECM stiffness to enable cells to sense each other over longer distances and thus coordinate their alignment. 
}


\subsubsection{Tissue morphogenesis}

\begin{figure}
  \includegraphics[width=\textwidth]{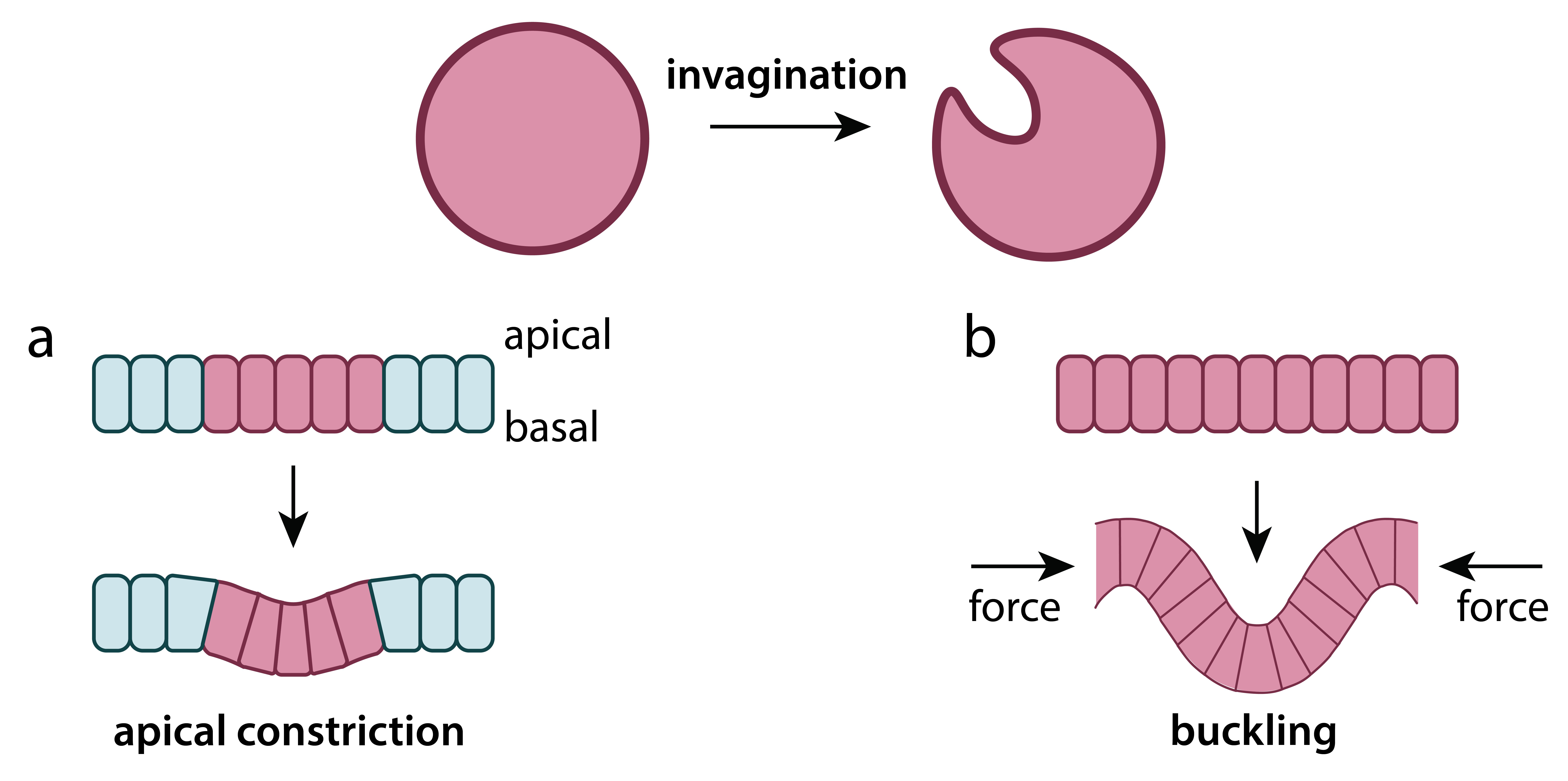}
\caption{\rs{Schematic depiction of processes of invagination. (a) Apical constriction: a cell shape change whereby the apical side of the cell actively contracts and shrinks, forming a concave indentation \citep{Martin_2014}. (b) Buckling: external mechanical forces cause a tissue sheet to bend, cell shapes change passively as a consequence of the change in tissue curvature \citep{Nelson_2016}.}}
    \label{fig:invag}
\end{figure}

\textbf{Cell-driven morphogenesis}.
Morphogenetic events can often be traced back to cellular changes, such as cell shape changes and cell rearrangements.
A textbook example is invagination, which is often ascribed to apical cell constriction (\rs{Figure \ref{fig:invag}}).
\cite{davidson1995sea} develops a highly detailed continuum mechanical model of invagination in the early sea urchin gastrula that includes separate layers for apical, lateral, and basal cell sides, as well as various layers of the ECM.
They then systematically test five competing hypotheses: apical cell constriction, lateral cell tractoring, apical ring contraction, apicobasal contraction, and ECM swelling. Surprisingly, all of these mechanisms are found to reproduce invagination, but require different mechanical parameters. 
For example, the model predicts that apical cell constriction requires similar cell and ECM stiffnesses, while an ECM swelling hypotheses requires the ECM to be much stiffer than the cells.

Cell rearrangement is another paradigmatic morphogenetic event, and an excellent use-case for ABMs.
\cite{longo2004multicellular} develop a cellular automaton model of radial intercalation in tissue extension of the blastocoel roof of \textit{Xenopus laevis} embryos. The model consists of several stacked cell layers, where at the very bottom they are lined with a layer of fibronectin ECM. Using phenomenological rules for cell motility and fibronectin deposition, they recapitulate the pattern of tissue elongation and basal fibronectin enrichment. 
The model predicts that groups of initially adjacent cells disperse laterally during intercalation, which they subsequently verify with cell transplantation experiments. 
The model also shows that the fibronectin layer reduces cell dispersion by promoting cell adhesion, and thus prevents overextension of the tissue.

\textbf{Somitogenesis}.
%
During vertebrate development, somites form by segmental epithelialization of mesenchymal cells derived from the presomitic mesoderm at the posterior end of the embryo. The cells in each segment become patterned by spatio-temporal gene activity oscillations in a mechanism known as the clock-and-wavefront model \citep{baker2006clock}.
\cite{truskinovsky2014mechanical} argues that chemical models of somitogenesis are insufficient to explain physical separation of the tissue into segments, and propose a mechanical model where inter-tissue coupling via fibronectin ECM plays a crucial role. This model predicts that segments form because the presomitic mesoderm is a relatively soft tissue that extends while mechanically connected to and confined by relatively rigid layers. 
A similar physical mechanism is used by \cite{nelemans2020somite} where the ECM is represented as a mesenchymal cell type in the context of a CPM. This work predicts that imposed exterior mechanical strain leads to the separation of somites, which is also experimentally verified by mechanical manipulation of tissue explants.
Supportive of the predictions of these modeling approaches, new intriguing experimental evidence indicates that the fibronectin ECM plays a role in regulating left-right somite symmetry \citep{naganathan2022left} and inter-tissue growth coordination \citep{guillon2020fibronectin}. 

\textbf{Folding morphogenesis.}
A recurring theme in morphogenesis is the reshaping of flat or tubular tissue into more complex topologies with folds, bends, and loops. Tissue folding can occur via multiple mechanisms, including cell growth, migration either of single cells or epithelial sheets, confinement-imposed forces by adjacent tissues, and also from a mechanical mismatch between adjacent tissue layers such as epithelia and ECM-rich mesenchyme \citep{hughes2018engineered, agarwal2022directed}. 
%
For example, \cite{Gardiner_2015} exert compressive forces on a subcellular ABM where cells and the BM are represented by clusters of particles. 
When lateral cell-cell adhesion is high, the tissue responds by mechanical buckling. 
In contrast, when vertical cell-ECM adhesion prevails, the tissue resists buckling and instead adapts to compression by cell shape changes to columnar morphology.
%
Although tissue folding is a highly reproducible process in embryonic morphogenesis, mimicking it \textit{in vitro} remains a challenge. 
To address this, \cite{hughes2018engineered} developed an experimental bioengineered tissue scaffold whose folding pattern is predicted using a continuum mechanical model. 
By combining experimental work with modeling, cell contraction is found to condense ECM into specific locations, which allows the tissue to locally support greater tensile strength. 
These bands of condensed ECM are thus akin to predetermined faultlines that facilitate tissue folding in a robust, predictable pattern.

An example of the flat to folded transition is seen in the \textit{Drosophila} wing imaginal disc - a bilayered epithelium surrounded by BM that develops stereotypical folds and eventually gives rise to the adult wing of the fly, \citep{martin2009cell}. Using a continuum mechanical model, \cite{keller2018influence} estimate mechanical properties of the tissue through comparison with stretching experiments. Interestingly, a stiffer layer of ECM at one side of the tissue is necessary to replicate experimental measurements.
\cite{tozluoglu2019planar} develop a continuum mechanics model of the wing disc and the underlying BM. 
In their model, folding at the stereotypic positions requires both differential planar tissue growth and mechanical constraining by uniform elastic compression from the BM. 
Though this model correctly predicts folding, it cannot resolve individual cells and the contribution of cell shape changes.
To address this, \cite{nematbakhsh2020epithelial} develop a 2D subcellular element model that includes detailed cell shape, nuclear shape and position, and homotypic and heterotypic cell-cell and cell-ECM contacts. Using a combination of modeling and experiments, they find that actomyosin contractility is required to initiate tissue folding, but is dispensable afterwards. 
In contrast, ECM tension cannot create the initial fold, but is necessary to maintain it after actomyosin activity ceases.

Several embryonic tissues start as a straight tube-like shape, then fold and twist to produce loops. 
A classic example is the vertebrate heart which undergoes cardiac looping, \citep{taber2003biophysical}. The embryonic heart consists of two epithelial cell layers enclosing a thick layer of ECM, the cardiac jelly. 
\cite{ramasubramanian2006computational} develop a continuum mechanics model of both the outer myoepithelium cell layer and the cardiac jelly. 
Each layer's material properties are adapted to match experimental data, including deformation terms to account for cell migration, cell growth, active contraction, and ECM swelling. 
This model predicts that swelling of the cardiac jelly, though not responsible for organ shape change, enhances structural integrity of the epithelial layer by balancing cell forces.

\begin{figure}
  \includegraphics[width=\textwidth]{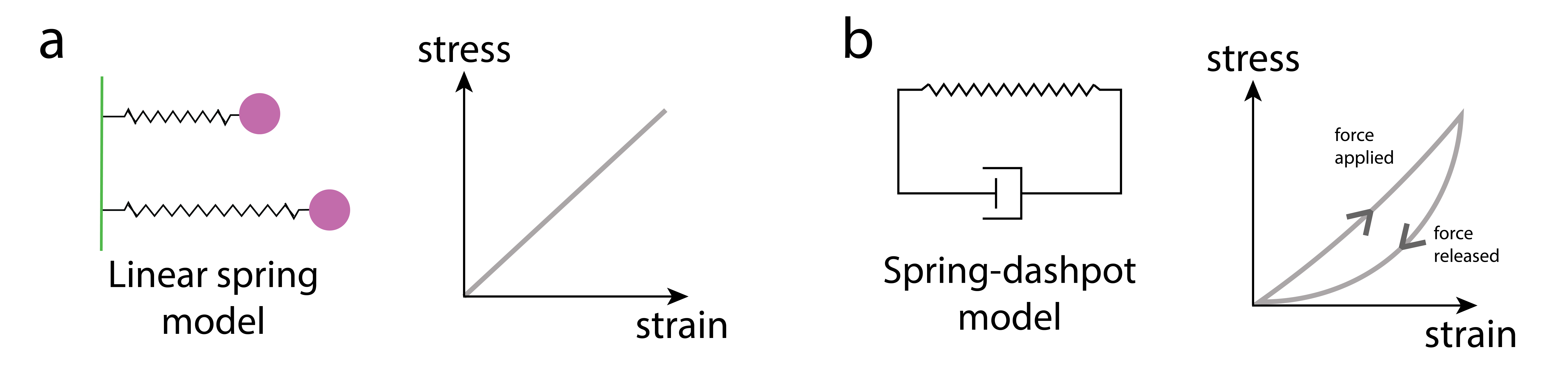}
\caption{\rs{Schematic of material models. (a) A spring-like model where the strain is linearly proportional to stress, and (b) a spring-dashpot model that depicts the stress-strain relationship in viscoelastic materials. The spring permits fast, unrestricted motion, whilst the  dashpot is effectively a damper that slows motion in the opposite direction.}}
    \label{fig:mech}
\end{figure}

%
The vertebrate intestine is another example of a tubular organ that bends and loops. 
The first step of gut looping is a leftward tilt in the tissue, which \cite{kurpios2008direction} model by representing cells as a collection of six membrane nodes and one internal node, all mechanically connected via \rs{spring-dashpot mechanics (Figure \ref{fig:mech}b}). 
The ECM is represented indirectly by a spring-like forces between adjacent cells. 
The model shows that, while cell shape change leads to a slight leftward tilt, it fails to explain the full range of motion of the tissue. 
Instead, the model predicts a synergistic effect of asymmetric swelling of the ECM combined with asymmetric cell shape changes.
Looking at the scale of the entire organ, \cite{savin2011growth} elegantly combine \textit{ex vivo} tissue explants with physical rubber models and computational discrete mechanical models to explain gut looping. 
Using a similar mechanical model as \cite{kurpios2008direction}, \cite{savin2011growth} show that looping emerges from a mismatch of the mechanical parameters of two adjacent and connected tissue layers, the intestinal epithelium and the ECM-rich mesentery. 
Much like in a bimetallic strip, differential expansion or shrinkage of one tissue leads to wrinkling, which manifests as looping in the tube-shaped tissue.

On a smaller spatial scale, the intestinal tissue starts as a flat epithelium on top of a BM that folds to form multiple undulations -- the bottom of these undulations will form the intestinal villi and crypts that house stem cells.
\cite{hannezo2011instabilities} develop a mechanical continuum model of the intestinal tissue, where epithelium, BM, and stroma are each modeled as layers with separate mechanical parametrization. They predict that different patterns can emerge by mechanical buckling -- spots, checkerboards, herringbone-stripes, and labyrinthine stripes -- and depend on parameter values for coupling of cell division and ECM curvature, and the parameters for pressure exerted by the intestinal epithelium. 
The elasticity of the BM is predicted to play a crucial role in stabilizing the amplitude of the villus folds.
%
\cite{dunn2012two} develop an ABM of the BM to study how intestinal crypts attain their morphology. 
They model the BM as discrete regions between epithelium and stroma, each with a preferred curvature maintained via \rs{spring-like forces (Figure \ref{fig:mech}a}). 
Starting from a flat sheet, a gradual change in BM local preferred curvature is imposed, which results in gradual invagination of the intestinal crypt. 
The depth of invagination also depends on the density of epithelial cells, which results from a combination of proliferation, death, and intercellular forces. 
For example, when cells adhere more strongly they become denser, and their collective force overcomes the ECM force and thereby reduces crypt depth.
The model thus predicts that deeper crypts emerge when intercellular connections are weak.
\cite{shyer2013villification} develop a more detailed continuum mechanics model of a section of the gut tube to model villus morphogenesis. 
They take into account not only the epidermal and mesenchymal layers (which includes the BM), but also surrounding muscular layers, each modeled as a separate material. 
This study elegantly demonstrates how mechanical coupling of layers growing or shrinking at different rates can lead to development of tissue folds. 
Interestingly, their model also recovers the same sort of patterns found by \cite{hannezo2011instabilities} in their more abstract model.

Another tube-like structure that becomes looped is the \textit{C. elegans} gonad, which consists of an epithelium ensheathed by a BM. 
\cite{agarwal2022directed} model the gonad as a rectangular structure with elastic walls representing the BM, and include cell-ECM adhesions using elastic springs. 
To test their hypothesis that asymmetric BM degradation induces looping, they impose different spring constants on one side of the tissue. 
This mechanical imbalance directs the tissue to always bend in the direction opposite to weaker adhesions, which the authors verify experimentally.

%

%
Many epithelial tissues such as lungs and secretory glands develop by a branching of initially symmetric epithelial structures. As in gut morphogenesis, these epithelia are usually ensheathed by a BM, and further surrounded by an ECM-rich mesenchymal tissue with contractile fibroblasts \citep{varner2015mechanically}. 
The interplay between epithelium and mesenchyme plays a central role in determining the position of branch points, but the nature of this interplay -- chemical or mechanical -- is under debate \citep{varner2015mechanically}.  
In an exploratory model, \cite{wan2008mechanics} investigate how mesenchymal fibroblast contraction may deform an epithelial bud at the onset of branching. 
Their mechanical continuum model explicitly includes both fibroblasts and ECM in the mesenchyme using a fluid mixture model. 
Variations in density and spatial distribution of contractile fibroblasts give rise to different morphologies in the epithelial bud such as dimples and clefts. 
%
\cite{varner2015mechanically} combine experimental work in lung bud explants with a continuum mechanical model of a growing flat epithelial sheet surrounded by viscoelastic medium. 
Their model predicts that explants branch due to growth in a confined environment, which they verify experimentally by embedding explants in increasing concentrations of Matrigel, a commercially available matrix.
Notably, the mechanism of shape generation is highly reminiscent of the models of gut and gonad morphogenesis discussed earlier, and suggest an elegant underlying principle of tissue morphogenesis: adjacent tissues each with different mechanical properties and different growth rates. 
Crucially, both of these properties are tunable by cells -- tissue mechanics by changing ECM composition, and growth by cell proliferation or ECM swelling. 
In essence, this suggests that ECM composition is part of evolution's toolbox in shaping the diversity of forms in the animal kingdom.

\section{Discussion}
The aim of this review is to highlight the variety of mathematical models and computational tools available to study the composition and function of the ECM across a range of biological systems and scales\rs{, and to direct the interested reader to relevant papers}. 
From the vast number of examples presented, it is clear that mathematical and computational \rs{investigations} can help illuminate key behaviors and mechanisms of \rs{interaction between cells and} the ECM. 

\rs{
Throughout this review, we highlight models of cell migration, including cancer cell progression and wound healing, alongside models of tissue structure and morphology, that concern the integrity and composition of a tissue.
Notably, ECM structure, particularly the mechanical properties, play a key role in both cell migration and tissue morphology. 
A prime example of how matrix stiffness affects cell migration is durotaxis, which different modeling approaches demonstrate requires force-dependent focal adhesion dynamics and contractile cytoskeletal activity \citep{rens2020cell, paukner2023key}. 
ECM stiffness also drives vasculature formation, with simulations from \cite{carrasco2023silico} and \cite{noerr2023optimal} highlighting its importance in coordinating cell behaviors via mechanical feedback. 
Fiber alignment and orientation play a central role in several models, including in migration and pattern formation. One such model of wound healing \citep{cumming2010mathematical} produces simulations concurrent with experimental data, emphasizing the importance and role of ECM fiber properties during scar formation. 
}

\rs{
A major benefit of the use of mathematical models in biology is their predictive power in guiding further experiments. Often, hypotheses generated \textit{in silico} are used to predict key parameters or mechanisms within a system, which in turn expedite corresponding experimental investigations by reducing their required scope. For example, the ABM for NCC migration considered by \cite{mclennan2022colec12} was used to predict that Trail and Colec12, factors expressed adjacent to the regions through which cranial NCCs migrate, must only be expressed for less than half of the length of the migratory domain for the confinement of migrating NCCs. This hypothesis was then tested in chick, where Colec12 and Trail were also found to be highly expressed for around one third of the length of the domain through which migration occurs. In the context of neural crest biology, a similar ABM was also used to predict distinct phenotypes in cranial neural crest cells at the leading and trailing edge of collectives. This prediction motivated a gene expression profile analysis of chick cranial neural crest cells, which also found differential gene expression across collectives \cite{mclennan2012multiscale}. Such findings highlight the predictive power of mathematical and computational modeling in helping to guide experiments towards mechanisms and parameters of interest in understanding observations \textit{in vivo}.

Another key purpose of mathematical models in biology is in elucidating unexplained processes, for example, the role of the ECM remodeling enzyme LOX during cancer cell migration \citep{Edalgo2018}. 
This particular model provides predictions that could guide therapeutic treatments, such as inhibiting LOX production to slow down ECM remodeling, thus preventing the spread of cancer. 
Another example is the investigation of how the ECM regulates the underpinning molecular mechanisms of cell migration \citep{Park_2017}. 
In understanding how ECM structure impacts the molecular mechanisms of cells, greater insight is developed into the design of targeted pharmaceutical treatments that can inhibit or express more migratory cell phenotypes.}

\rs{Nevertheless, in order to refine and validate these models, it is crucial to couple modeling} with experimental investigations. 
To facilitate such collaborative efforts, in the following discussion, we aim to provide brief guidance for determining the most appropriate modeling framework for a given study. 
We discuss the challenges faced when pairing mathematical and computational models with sparse experimental data and also the benefits of fostering interdisciplinary collaborations between theoreticians and experimentalists.

\subsection{Selecting an appropriate modeling framework}

Many mathematical models of biological processes neglect the structure of the ECM and its interactions with cells, and this review serves to highlight modeling efforts that explicitly consider the ECM and its role in processes such as tumor progression (Section \ref{cancer}), wound healing (Section \ref{woundHealing}), and tissue morphogenesis (Section \ref{morphogenesis}). The most appropriate modeling framework to adopt depends on various factors including the complexity of the system at hand, the characteristic time and length scales of relevant processes, and the overall purpose of a study. In this section, we highlight key considerations for selecting a model framework and list the relative advantages and disadvantages of each framework considered. 

When considering the composition of the ECM itself, the most suitable modeling approach depends primarily on the spatial scale at which the ECM is to be studied. For example, at the tissue level, simulating each constituent cell and ECM particle will incur significant computational costs \citep{Stack2022, saxena_biofvm-x_2021} that make repeated simulations or parameter sweeping infeasible. 
Therefore, over large length scales, continuum representations of the ECM are often favored over ABMs due to their lower computational demands. 
The coarse-graining of ABMs offers a potential compromise between microscopic detail and macroscopic simplicity \citep{penington2011building, simpson2009diffusing}. 
For example, work by \cite{Voul2017} \rs{derives a continuum representation of an ABM for tumor cell reprogramming, which is found to produce similar results to simulations of the underlying ABM for certain parameter regimes}.
However, an advantage of using purely agent-based representations of the ECM is that they permit greater interpretability than continuum models through their \rs{detailed} descriptions of cell behaviors and cell-cell communication. 
Furthermore, ABMs allow an even greater emphasis to be placed on behaviors observed at the cell-scale and can often capture interactions between cells and the ECM that continuum frameworks cannot. For example, \cite{Bull2020} develop a hybrid ABM model that is required to capture microbead infiltration patterns in tumor spheroids.
For subcellular level studies, discrete mechanical models may be most suitable, as they allow a more detailed representation of cell morphology and cell-cell interactions based on the underlying physical properties of cells. 
However, the use of such models is also primarily limited by the computational resources required to simulate many cells at a subcellular resolution.
Alternatively, when a system of interest comprises elements at vastly different characteristic length-scales, combining agent-based and continuum approaches, for example, into a hybrid model may be the most desirable solution in that it offers accuracy of each relative counterpart in the system, without the computational cost of a fully agent-based system. 
It is, therefore, clear that the scale of the tissue and the scale at which the interactions of interest are occurring is vital in establishing the most appropriate modeling framework. 


As well as defining the spatial scale of the system, it is also important to consider the availability of experimental data for a computational investigation. Mechanical models require precise material properties and geometrical data to be accurately calibrated, whereas continuum models require only macroscopic data, such as spatially averaged mechanical properties and concentrations of ECM constituents. 
In many cases, obtaining biomechanical measurements requires specialized expertise that may not be available to most experimental researchers. In contrast, ABMs necessitate granular data at the cellular or molecular level, such as individual cell behaviors and interaction rules, that are ideal when high-resolution experimental data is abundant. Hybrid models, such as coupled continuum and mechanical models, are versatile and can be adapted to the available data, making them suitable for scenarios with variable data resolutions.
The choice among these frameworks should, therefore, be closely aligned with the quality, quantity, and scale of the available data, ensuring that the chosen model can be adequately parameterized and validated to produce reliable insights into ECM composition and its effects on biological processes.

Using data generated in experiments, a common paradigm in biological modeling is to iteratively refine models upon comparing model predictions with corresponding \textit{in vitro} and \textit{in vivo} studies. As such, another important consideration in selecting a model is the extent to which it may be adapted and refined to account for new experimental findings. Continuum models are inherently adaptable at the macroscopic level, easily integrating changes in material properties and boundary conditions, which makes them suitable for a broad range of tissue-scale scenarios of varying degrees of complexity. 
Conversely, ABMs are easily mechanistically interpretable and as a result, are particularly well suited to refinement at an individual cell/ECM constituent level by permitting biological mechanisms to be built into models with relative ease. 
Furthermore, ABMs often contain more parameters that can be independently varied for comparison to wide-ranging biological phenomena. In general, mechanical models are less flexible in terms of biological adaptability but can be highly amenable to incorporating new mechanical data or in simulating different physical scenarios. 
The flexibility of mechanical models lies in their capacity to adjust to new geometries, loading conditions, and material properties, all of which are crucial when exploring the bio-mechanical aspects of the ECM.

Finally, the complexity of the biological system often dictates the choice of mathematical model and resulting number of parameters in the system, which represent quantities such as exchange rates, substrate concentrations, and forces. In general, it is best to consider the number of parameters attainable via experimental exploration when choosing the most suitable modeling framework. It is also possible to determine unknown parameter values through model fitting. With this in mind, the larger the number of parameters, the harder it can be to fit the results of the model to \textit{in vivo} and \textit{in silico} data. 
Moreover, some of these parameters may be difficult to interpret biologically, and may in fact be an artifact of the model development, for example, ABMs often include a parameter that sets a threshold distance to limit cell-cell interactions.
 
In cases where fitting the model to experimental data is not possible, other approaches can be used, such as machine learning algorithms \citep{preen_towards_2019}, that explore the parameter space to find the optimum values and reach an objective function \citep{ozik_high-throughput_2018}. \rs{These, however, have the drawback of requiring a large amount of data, upwards of ten times the number of samples versus number of features or parameters. Additionally, interpretation of these models can be very difficult, with the `black box' nature of many of these models providing metrics for the predictive accuracy of a model, but inhibiting interpretation of the underlying biological meaning \citep{Xu_2019}.}

In summary, carefully assessing and balancing elements such as the scale of the biological system, the specific objectives of the study, and the accessibility and type of relevant data, is crucial when considering the most suitable modeling framework for the biological system of interest.

\rs{\subsection{Open Challenges}

Due to the inherent complexity of the ECM, containing components that vary vastly in composition and size, it can be difficult to quantify its properties experimentally. Methods mentioned in Section \ref{sec:experiment} can identify ECM components and characterize ECM structure, chemical composition, and mechanics. However, the specialized expertise essential to each method underscores the importance of interdisciplinary collaborations. Advances in obtaining quantitive and qualitative data of the ECM will greatly enhance \textit{in silico} models by allowing for more accurate predictions to be made through enhanced model parameterization.

One of the most pertinent aspects of mathematical modeling in biology is the integration of models over multiple spatial and temporal scales. 
For example, molecular interactions occur at a vastly different rate from cellular or tissue level dynamics and therefore must be modeled at a different temporal scale. 
Linking such scales can be difficult depending on the numerical approaches taken, where the simulations and results from one scale define the parameters and variables at another. Multi-scale models are an excellent tool for simulating processes that occur over vastly different time or length-scales, but can become prohibitively complex as more interactions and biological phenomena are considered. The development of a multi-scale model most certainly means an increase in the size of the model and computational power needed for simulations. As a model increases in size, typically the number of variables and parameters will also increase. Some of these issues can be tackled, for example if there are sufficient data available, such that unknown parameters can be constrained to a realistic regime. However, without such constraints, these models can become too complex to systematically analyze and biological interpretation can be lost. 
 
Computational implementation can also be a barrier to developing such models. Although there are many available software packages for building computational models (see Table \ref{tab:resources} for a list of resources), they are built with a specific modeling approach in mind. Thus, often one needs to develop new codes tailored to a specific biological question, reducing the possibility of fully leveraging hybrid or multi-scale systems, and requiring extensive programming ability. 

All of these challenges apply to models that either do or do not include the ECM. Models that include the ECM are intrinsically more complex due to at least one additional variable in the model (or more if including specific components of the ECM) and their associated properties. The implementation of large, all-encompassing models is something many aspire to, and work is ongoing in the field of mathematical biology to improve and perfect these models to elucidate the most relevant and important information \citep{Eftimie_2022}. 
}

\subsection{Mathematical modeling as a tool in collaborative biological research}
This review highlights that the development of models and the interpretation of wet lab results are intricately linked, drawing insights and questions from the existing literature and data. 
In general, mathematical models are developed to investigate previously unexplored experimental conditions \rs{or to predict behavior in conditions that are experimentally intractable}. Hence, mathematical models provide an alternative way to explore new theories and hypotheses that can then be tested and validated experimentally. 

A major benefit of using mathematical models to study biological systems is their ability to provide an efficient approach to rapidly executing numerous simulations with wide-ranging parameter sets, such that relevant biological parameter ranges may be quickly identified for further testing in experiments. 
This route also proves to be faster and vastly less expensive than the cost and time-intensive nature associated with setting up and performing experiments both \textit{in vitro} and \textit{in vivo}. 
Instead, models excel in exploring conditions that, in the lab, are challenging to replicate or analyze with available equipment. 
\rs{In the development of novel therapies, models may be used to study the effects of biological perturbations in a system. This, in turn, can be used to predict or highlight parameters of particular importance in the prevention of diseases, eliminating the need for costly interventions after the onset of pathogenesis.}
\rs{Furthermore, it is often difficult, or impossible, to experimentally stain numerous cell phenotypes for identification in microscopy. In such cases, mathematical models can be used to predict the presence of certain phenotypes that experiments cannot replicate, in order to enhance the understanding of the underlying cell-ECM interactions, and help develop targeted treatments specific to a given phenotype.}
Additionally, mathematical simulations provide a systematic approach to exploring \rs{mechanical} perturbations within systems, for example, stiffness of the ECM \citep{Yuste2022}, and structure of fiber networks and their alignment \citep{lee2017local}.   

In highly complex systems where designing experiments and gathering data can be difficult, model predictions aim to help in determining significant molecules or pathways in a system. 
This approach simplifies the system and narrows down the scope of focus towards interesting processes and behaviors, helping to guide experimental design.
 
In this way, mathematical models work most effectively when there is direct synergy between both the experimental and computational counterparts. As such, collaborative efforts often unveil new research questions during the construction and validation of biological hypotheses, establishing a continuous cycle of model development, refinement, and experimental data generation that yields new insights and perspectives.

\rs{\subsection{Tables}}

\begin{table}[ht]
\centering
\begin{tabular}{ p{4.5cm} | p{7.5cm} | p{4.5cm} }
\hline \Tstrut
ECM component & Function & ECM Location \\ [0.5ex]
 \hline \Tstrut
\textbf{Proteins} & & \\ [0.5ex]
Fibril collagens, e.g., type I, II, III & Provides structural integrity and strength to the tissue. & Interstitial ECM  \\ [0.5ex]
Network-forming collagen, e.g., type IV & Sheet-like ECM forms from fibers. & Basement membrane  \\ [0.5ex] 
Elastin & Provides tissue with elasticity and load-bearing capabilities. & Interstitial ECM  \\ [0.5ex]
\textbf{Glycoproteins} & &  \\ [0.5ex]
Fibronectin & Cross-links with ECM proteins, influences cell adhesion and migration. & Interstitial ECM (also found concentrated near basement membrane)  \\ [0.5ex] 
Laminin & Forms a mesh-like network to aid in cell adhesion and migration. & Basement membrane  \\ [0.5ex]
Nidogen & Promotes BM structural integrity by facilitating laminin and collagen type IV cross-linking. & Basement membrane  \\ [0.5ex]
Proteoglycans & Regulate ECM structure, e.g., perlecan, performing a similar role to nidogen, and supports cellular functions by storing growth factors and other signaling molecules. & Interstitial ECM and Basement membrane  \\ [0.5ex]
\textbf{Enzymes} & &   \\ [0.5ex]
Lysyl oxidase (LOX) & Mediates ECM structure, facilitating ECM protein cross-linking. & Interstitial ECM  \\ [0.5ex]
Matrix metalloproteinases (MMPs) & Degrades ECM proteins such as collagens, fibronectin and laminin. & Interstitial ECM and Basement membrane  \\
\hline 
\end{tabular}
\caption{\rs{Table describing some of the major ECM components alongside their functions and locations.}}
\label{tab:ecmfunc}
\end{table}

\begin{table}[ht]
\centering
\begin{tabular}{p{3cm} | p{14cm}}
\hline \Tstrut
Term & Definition \\ [0.5ex]
\hline \Tstrut
ABM & A mathematical and computational model comprised solely of multiple interacting individual elements. \\ [0.5ex]
Apical surface & Surface of an epithelial cell that faces the lumen, or inside of a tubular structure, such as vessels or intestine. \\ [0.5ex]
Basal surface & Surface of an epithelial cell that adjacent to the basement membrane and underlying tissue. \\ [0.5ex]
Cell tractoring & A process in which the cell cortex creates a track around the cell to move the cell relative to its neighbors. \\ [0.5ex]
Chemotaxis & The movement of cells in response to some chemical stimulus, for example, chemoattractants. \\ [0.5ex]
Durotaxis &  The movement of cells in response to a gradient of extracellular stiffness. \\ [0.5ex]
Haptotaxis & The movement of cells in response to a gradient of adhesive substrates, for example, integrins.  \\ [0.5ex]
Intercalation & A process in which neighboring cells exchange places. \\ [0.5ex]
Invagination & The process in which a surface folds back on itself, forming a cavity or pouch. \\ [0.5ex]
ODE & A differential equation in one variable, for example, time. \\ [0.5ex]
Parameterization & A mathematical process used to express a system through a function of parameters. \\ [0.5ex]
PDE & A differential equation in more than one variable, for example, time and space. \\ [0.5ex]
Septa & A dividing wall or membrane between tissues. \\ [0.5ex]
\hline 
\end{tabular}
\caption{\rs{Glossary of important terms.}}
\label{tab:glossary}
\end{table}

\begin{table}[ht]
\centering
\begin{tabular}{l | l}
\hline \Tstrut
Abbreviation & Definition \\ [0.5ex]
\hline \Tstrut
ABM & Agent-based model \\ [0.5ex]
BM & Basement membrane \\ [0.5ex]
CPM & Cellular Potts model \\ [0.5ex]
ECM & Extracellular matrix \\ [0.5ex]
EMT & Epithelial-to-mesenchymal transition \\ [0.5ex]
FEM & Finite element method \\ [0.5ex]
LOX & Lysyl oxidase \\ [0.5ex]
MDE & Matrix-degrading enzyme \\ [0.5ex]
MMP & Matrix metalloproteinases \\ [0.5ex]
NCC & Neural crest cell \\ [0.5ex]
ODE & Ordinary differential equation \\ [0.5ex]
PDE & Partial differential equation \\ [0.5ex]
VEGF & Vascular endothelial growth factor \\ [0.5ex]
\hline 
\end{tabular}
\caption{\rs{Table of abbreviations and their definitions.}}
\label{tab:abbrev}
\end{table}

\begin{table}[!htbp]
\centering
\begin{tabular}{p{3.5cm} | p{9cm} | p{4cm}}
\hline \Tstrut
Resource & Description & Publications \\ [0.5ex]
\hline \Tstrut
CompuCell3D \url{https://compucell3d.org/} & A C++ and Python-based software environment coupling ABMs for cellular processes to reaction-diffusion PDE models for chemicals in the cellular microenvironment. & \cite{Swat_CC3D, izaguirre_c_2004}
\\                                                                                                                                                                   
Chaste \vspace{-0.1cm}  \newline \url{https://www.cs.ox.ac.uk/chaste/index.html} & A C++ based open-source environment for the simulation of multi-scale models of cellular processes coupling discrete lattice-based and lattice-free models for cell populations to continuum models for chemical transport.  & \cite{PITTFRANCIS_2009, Mirams_2013,Cooper_2020}                                                                                                                                                                                                          \\ 
CellSys \vspace{-0.1cm} \newline \url{https://bio.tools/cellsys} & A C++ based modular software tool for off-lattice simulations of 2D and 3D growth and morphogenesis processes, permitting real-time 3D visualization of simulations. &  \cite{Hoehme_2010}                                                                                                                                                                                            \\
PhysiCell \vspace{-0.1cm} \newline \url{http://physicell.org/} & A C++ based open-source framework for 3D physics-based simulations of multicellular systems. An agent-based framework for cell movement and interactions is coupled to a PDE solver for chemicals secreted by cells. & \cite{Ghaffarizadeh_Heiland_Friedman_Mumenthaler_Macklin_2018}
 \\  
PhysiBoss \vspace{-0.1cm}\newline \url{https://github.com/PhysiBoSS} & A PhysiCell-based framework that allows for the simulation of signaling and regulatory networks in individual cells. & \cite{letort_physiboss_2019, ponce-de-leon_physiboss_2023}                                                                                                                                                                                                                           \\
Morpheus\vspace{-0.1cm} \newline \url{https://morpheus.gitlab.io} & An open-source multi-scale modeling environment for the simulation of cell-based models coupling 2D and 3D cellular Potts models, ODEs, and PDEs.  & \cite{Starruss_2014}                                                                                                                                                                                                                                                                                                                                                                                                                                                                                                                                                              \\
VirtualLeaf \vspace{-0.1cm}\newline \url{https://code.google.com/archive/p/virtualleaf/} & An open source C++ based ABM software to model plant cells and tissues. & \cite{Merks_2011, Merks_2012}                                                                                                                                                                                                                                                                        \\
Tissue Forge \vspace{-0.1cm}\newline \url{https://compucell3d.org/TissueForge} & A C, C++, and Python-based interactive environment for the simulation of biological and biophysical systems from sub-cellular to  tissue-level scales.  &  \cite{Sego_2023}
\\    
COMSOL \vspace{-0.1cm}\newline \url{https://www.comsol.com} & An FEM-based numerical PDE solver environment for the simulation of complex biological and biophysical systems.  & \cite{multiphysics1998}                                                                                                                                                                                         \\
FEniCS \vspace{-0.1cm}\newline \url{https://fenicsproject.org} & An open-source platform for solving complex PDEs with an FEM, offering C++ and Python interfacing.   & \cite{Alnaes_2015}
\\   
VCell \vspace{-0.1cm}\newline \url{https://vcell.org}  & A Java, C++, and Perl-based open-source platform for the simulation of biochemical and electrophysiological systems through deterministic and stochastic ODE- and PDE-based models with data-integration capabilities.  &  \cite{Schaff_1997,Cowan_2012}  \\
\hline                                                                                                                           
\end{tabular}
\caption{\rs{A list of computational resources for modeling cell-ECM interactions in cell migration and tissue morphogenesis.}}
\label{tab:resources}
\end{table}


\section*{Conflict of Interest Statement}

\rs{Author MBe is employed by Optics11 Life Inc. The remaining authors declare that the research was conducted in the absence of any commercial or financial relationships that could be construed as a potential conflict of interest.}

\section*{Author Contributions}
RMC, SJ and ET share first authorship of the manuscript.

\noindent RMC: Conceptualization, Investigation, Writing - original draft and review \& editing;
SJ: Conceptualization, Investigation, Visualization, Writing - original draft and review \& editing;
ET: Conceptualization, Investigation, Writing - original draft and review \& editing;
ZB: Investigation, Writing - original draft;
MBe: Investigation, Visualization, Writing - original draft;
MBo: Investigation, Writing - original draft;
QJSB: Writing - original draft and review \& editing;
JM: Conceptualization, Writing - original draft and review \& editing;
MR: Investigation, Writing - original draft and review \& editing;
YY: Investigation, Writing - original draft and review \& editing;
RS: Conceptualization, Investigation, Visualization, Writing - original draft and review \& editing, Supervision.

\section*{Funding}
RMC is supported by funding from the Engineering and Physical Sciences Research Council (EP/T517811/1) and the Oxford-Wolfson-Marriott scholarship at Wolfson College, University of Oxford. 
SJ receives funding from the Biotechnology and Biological Sciences Research Council (BBSRC) (BB/T008784/1). 
ET is funded by the Dutch Research Council (NWO) grant number VI.Veni.222.323.
ZB is supported by funding from the Engineering and Physical Sciences Research Council (EP/T517914/1).
MBo acknowledges the University of Birmingham.
QJSB is funded by Dutch Research Council (NWO) by the ENW-XL program 'Active Matter Physics of Collective Metastasis' (OCENW.GROOT.2019.022). 
JM was supported in part by NSF Grant 1735095 - NRT: Interdisciplinary Training in Complex Networks and Systems, the Jayne Koskinas Ted Giovanis Foundation for Health and Policy, and the Breast Cancer Research Foundation.
MR is funded by PerMedCoE.
YY is funded by the Engineering and Physical Sciences Research Council (EP/W524311/1) and supported by the Travel for Research \& Study Grant at St Hilda’s College, University of Oxford.

\section*{Acknowledgments}
Authors would like to acknowledge Cicely Macnamara and Robyn Shuttleworth, alongside Centre Europ\'een de Calcul Atomique et Mol\'eculaire (CECAM) and the Lorentz Centre for hosting the workshop "The Extracellular Matrix: How to model structure complexity" in which this collaborative group was formed.  


\bibliographystyle{abbrvnat}
\bibliography{ECM2}
\end{document}